\documentclass[12pt,preprint]{aastex}

\shorttitle{Fast Solar Wind}
\shortauthors{van Ballegooijen \& Asgari-Targhi}

\begin{document}

\title{Direct and Inverse Cascades in the Acceleration Region \\
of the Fast Solar Wind}

\author{A. A. van Ballegooijen\altaffilmark{1},
M. Asgari-Targhi\altaffilmark{2}}
\altaffiltext{1}{5001 Riverwood Avenue, Sarasota, FL 34231, USA}
\altaffiltext{2}{Harvard-Smithsonian Center for Astrophysics, 
60 Garden Street, Cambridge, MA 02138, USA}

\begin{abstract}
Alfv\'{e}n waves are believed to play an important role in the heating and acceleration
of the fast solar wind emanating from coronal holes. Nonlinear interactions between the
dominant ${\bf z}_{+}$ waves and minority ${\bf z}_{-}$ waves have the potential to
transfer wave energy either to smaller perpendicular scales (``direct cascade") or to
larger scales (``inverse cascade"). In this paper we use reduced magnetohydrodynamic (RMHD)
simulations to investigate how the cascade rates $\epsilon_{\pm}$ depend on perpendicular
wavenumber and radial distance from Sun center. For models with a smooth background
atmosphere we find that an inverse cascade ($\epsilon_{+} < 0$) occurs for the dominant
waves at radii between 1.4 and 2.5 $R_\odot$ and dimensionless wavenumbers in the inertial
range ($15 < a_\perp < 44$), and a direct cascade ($\epsilon_{+} > 0$) occurs elsewhere.
For a model with density fluctuations there are multiple regions with inverse cascade.
In both cases the cascade rate $\epsilon_{+}$ varies significantly with perpendicular
wavenumber, indicating that the cacsade is a highly non-local process. As a result of the
inverse cascades, the enery dissipation rates are much lower than expected from
a phenomenological model, and are insufficient to maintain the temperature of the background
atmosphere. We conclude that RMHD models are unable to reproduce the observed properties
of the fast solar wind.
\end{abstract}

\keywords{Magnetohydrodynamics (MHD) - solar wind - Sun: corona - Sun: magnetic fields - 
turbulence - waves} 

\clearpage

\section{INTRODUCTION}

The fast solar wind emanating from coronal holes is believed to be driven by
Alfv\'{e}n waves that propagate outward along the open field lines
\citep[e.g.,][]{Parker1965, Heinemann1980, Velli1993, Dmitruk2002, Suzuki2005,
Cranmer2015}. In situ observations in the heliosphere indicate that the waves are
in a turbulent state with a broad spectrum of wavenumbers and frequencies
\citep[e.g.,][]{Coleman1968, Belcher1971, Hollweg1986, Matthaeus1990, Bale2005,
Borovsky2012}.
Alfv\'{e}n waves have also been detected by remote-sensing of the solar atmosphere
\citep[][]{Tomczyk2007, Tomczyk2009, DePontieu2007, Threlfall2013, Tian2011, Tian2014,
Morton2015, Liu2015}.
Alfv\'{e}n waves are a prime candidate for heating and accelerating the fast wind
because they have the ability to transport energy over large distances in the corona
\citep[][]{Barnes1966, Belcher1971a, Hollweg1973, Jacques1977, Velli1989, Marsch1997,
Matthaeus1999, Suzuki2006, Cranmer2007, Chandran2011}.
The slow solar wind may also be driven by Alfv\'{e}n waves \citep[][]{Oran2015}.

Turbulent cascade has long been considered a promising mechanism for dissipation of
Alfv\'{e}n waves in the solar wind \citep[e.g.,][]{Hollweg1982, Hollweg1986, Velli1989}.
Nonlinear interactions between counter-propagating Alfv\'{e}n waves are known to produce
turbulence \citep[][]{Iroshnikov1963, Kraichnan1965, Shebalin1983, Goldreich1995,
Goldreich1997, Bhattacharjee2001, Maron2001, Cho2002}. 
The turbulence can be described in terms of Elsasser variables, ${\bf z}_{\pm} =
{\bf v}_1 \mp {\bf B}_1 / \sqrt{4 \pi \rho_0}$, where ${\bf B}_1$ and ${\bf v}_1$ are
the magnetic- and velocity fluctuations of the waves, and $\rho_0$ is the mean plasma
density \citep[][]{Elsasser1950}. The ${\bf z}_{+}$ and ${\bf z}_{-}$ waves are
linearly coupled due to radial gradients in plasma density and magnetic field strength
\citep[][] {Heinemann1980, Velli1993, Hollweg2007}, a process often described as wave
``reflection". Hence, the dominant ${\bf z}_{+}$ waves produce a lower level of
${\bf z}_{-}$ waves, which we refer to as the ``minority" waves. In the solar wind
the dominant waves are outward-propagating, but the minority waves can have both inward-
and outward-propagating components \citep[e.g.,][]{Velli1989, Verdini2009, Perez2013}.
\citet[][]{Matthaeus1999} proposed that the corona may be heated by Alfv\'{e}n wave
turbulence driven by nonlinear interactions between the ${\bf z}_{+}$ and ${\bf z}_{-}$
waves. Detailed models of the solar wind based on these ideas since have been developed
\citep[e.g.,][]{Cranmer2005, Cranmer2007, Verdini2007, Verdini2010, Chandran2011,
Sokolov2013, Oran2013, vanderHolst2014, Lionello2014}. \citet[][]{Woolsey2015} have shown
that the turbulent heating varies strongly in time and space.
However, it should be kept in mind that turbulent cascade is not the only mechanism
for dissipating Alfv\'{e}n waves in the corona. Nonlinear coupling between Alfv\'{e}n-
and compressive waves may also play an important role \citep[][]{Kudoh1999, Moriyasu2004,
Suzuki2005, Suzuki2006, Matsumoto2010, Chandran2005, Cranmer2012}.

Models of solar wind turbulence have been developed by several authors. One approach is to
use the ``shell" model, which simplifies the nonlinear interactions by reducing the number of
wave modes that are allowed to interact \citep[][]{Velli1989, Buchlin2007}. This model has
the great advantange that very high perpendicular wavenumbers can be reached. \citet[][]
{Verdini2009} and \citet[][]{Verdini2012} used the shell model to study the formation and
evolution of a turbulent spectrum of Alfv\'{e}n waves produced by linear and nonlinear
wave couplings. Another approach to turbulence modeling is to perform direct numerical
simulations using the reduced magnetohydrodynamic (RMHD) equations \citep[e.g.,][]
{Dmitruk2002, Dmitruk2003, Perez2013}. These equations include the nonlinear couplings
that lead to turbulence, but omit other effects such as the coupling between Alfv\'{e}n-
and compressive waves. \citet[][]{Dmitruk2002} argued that reflection-driven turbulence
provides a robust heating mechanism that can explain the observed temperatures in the
region below 2 $R_\odot$. \citet[][]{Dmitruk2003} showed that a high dissipation efficiency
can be obtained when the nonlinear time scale of the turbulence is less than the Alfv\'{e}n
crossing time. \citet[][]{Perez2013} were the first to include the effects of the solar wind
flow on wave propagation in the RMHD model. They found that up to one third of the wave
energy launched at the coronal base is dissipated in the corona below the Alfv\'{e}n
critical point, and another third goes into doing work on the solar wind outflow.

In a previous paper \citep[][hereafter paper~I]{vanB2016} we presented RMHD simulations of
Alfv\'{e}n wave turbulence for the fast solar wind emanating from a polar coronal hole.
This modeling is an extension of our earlier work on turbulence in coronal loops
\citep[e.g.,][]{vanB2011, vanB2014, Asgari2012, Asgari2013, Asgari2014}.
Paper~I includes the effects of the solar wind flow
on Alfv\'{e}n-wave propagation. Two models of the solar wind are considered. In the first
model the plasma density and Alfv\'{e}n speed vary smoothly with height along the modeled
flux tube. We find that for this ``smooth" model the linear wave coupling is relatively
weak, producing only a low level of minority waves. Therefore, the energy dissipation rate
of the turbulence is insufficient to maintain the temperature of the background atmosphere.
We also present a second model with additional, random density variations that approximate
the effects of compressive MHD waves in the solar wind. We find
that such spatial variations in density can significantly enhance the minority waves and
thereby the turbulent dissipation rates.

The results of paper~I led us to conclude that interactions between Alfv\'{e}n- and
compressive waves may play an important role in the turbulent heating of the fast solar
wind. However, this conclusion is somewhat premature because the reason(s) for the low
dissipation rates are not yet well understood. In particular, we do not know whether
the low rates are a real physical effect or a numerical artifact. In the present paper
we further improve our numerical model, and we compute for the first time the cascade
rates $\epsilon_{\pm}$ as functions of perpendicular wavenumber and radial distance from
Sun center. Cascade rates have also been measured in the solar wind, using third-order
structure functions \citep[e.g.,][]{Stawarz2009, Coburn2014}. We find that analysis of
the cascade rates can shed light on the question why ``smooth" solar wind models produce
relatively low dissipation rates.

\section{CASCADE RATES IN REDUCED MHD TURBULENCE}
\label{sect:cascade}

The RMHD equations describe the nonlinear interactions between counter-propagating
Alfv\'{e}n waves \citep[e.g.,][]{Strauss1976, Strauss1997}. The waves are assumed to
propagate in a medium with a fixed background magnetic field ${\bf B}_0 ({\bf r})$
and density $\rho_0 ({\bf r})$, where ${\bf r}$ denotes the position.
Specifically, we consider a thin, open magnetic flux tube extending radially from the
Sun inside a coronal hole, so the field strength $B_0$ and density $\rho_0$ are
functions of radial distance $r$ from Sun center. The waves can be described in terms
of Elsasser variables, ${\bf z}_{\pm} (x,y,r,t) \equiv {\bf v}_1 \mp {\bf B}_1 /
\sqrt{4 \pi \rho_0}$, where ${\bf B}_1$ and ${\bf v}_1$ are the magnetic- and velocity
fluctuations of the waves \citep[][]{Elsasser1950}, $x$ and $y$ are the coordinates
perpendicular to the flux tube axis, and $t$ is the time.
\citet[][]{Oughton2001} and \citet[][]{Dmitruk2003} were the first to present RMHD
simulations for such an open flux tube, and found that reflection-driven turbulence can
be maintained in this environment despite the fact that the waves can escape into the
heliosphere. \citet[][]{Perez2013} included the effects of the solar wind outflow on
the waves, and presented a variety of RMHD models with different perpendicular correlation
lengths and correlation times of the imposed footpoint motions. In paper~I we
investigated whether RMHD models of wave turbulence can explain the observed heating
of the fast solar wind. In the present work we continue this investigation with a
more detailed analysis of the cascade processes. We shall refer to the ${\bf z}_{+}$
and ${\bf z}_{-}$ waves as the dominant and minority waves, respectively.

The Elsasser variables are nearly incompressible velocity fields, and can be written
as ${\bf z}_{\pm} = \nabla_\perp f_{\pm} \times \hat{\bf B}_0$, where $\nabla_\perp$
is the spatial derivative in the $x$ and $y$ directions, $f_{\pm} ({\bf r},t)$ are the
velocity stream functions, and $\hat{\bf B}_0 (x,y,r)$ is the unit vector along the
background field. The RMHD equations can be written as
\begin{eqnarray}
\frac{\partial \omega_{\pm}} {\partial t} & = & 
     - ( u_0 \pm v_{\rm A} ) \frac{\partial \omega_{\pm} } {\partial r} \nonumber \\
 & & + \frac{1}{2} \left( \frac{dv_{\rm A}} {dr} \pm \frac{u_0}{2 H_\rho} \right)
       \left( \omega_{+} - \omega_{-} \right) + \frac{u_0} {2 H_{\rm B}}
       \left( \omega_{+} + \omega_{-} \right)  \nonumber \\
 & & - \onehalf [ \omega_{+} , f_{-} ] - \onehalf [ \omega_{-} , f_{+} ] 
     \pm \nabla_\perp^2 \left( \onehalf [ f_{+} , f_{-} ] \right) , \label{eq:rmhd1}
\end{eqnarray}
where $\omega_{\pm} \equiv - \nabla_\perp^2 f_{\pm}$ are the vorticities associated with
the dominant and minority waves, $u_0 (r)$ is the outflow velocity of the wind,
$v_{\rm A} (r)$ is the Alfv\'{e}n speed, $H_B (r) \equiv B_0 / (d B_0 /dr)$ is the
magnetic scale length, and $H_\rho (r) \equiv \rho_0 / (d \rho_0 /dr)$ is the density scale
length. Equation (\ref{eq:rmhd1}) can be derived from the expressions given in
\citet[][]{Perez2013} and paper~I. The first term on the right-hand side of this equation
describes the effects of wave propagation, and the second and third terms describe the
linear couplings between the dominant and minority waves. The bracket operator
$[ \cdots , \cdots ]$ is defined by
\begin{equation}
b(x,y,r,t) = [f,g] \equiv \frac{\partial f} {\partial x} \frac{\partial g} {\partial y}
- \frac{\partial f} {\partial y} \frac{\partial g} {\partial x} ,
\label{eq:bracket1}
\end{equation}
where $f(x,y,r,t)$ and $g(x,y,r,t)$ are two arbitrary functions. All nonlinearities of
the RMHD model are contained within such bracket terms. In equation (\ref{eq:rmhd1}) we
omit the dissipative terms, which will be described in more detail below.

Most RMHD models use a spectral method in which all functions of $x$ and $y$ are
written in terms of a set of normalized basis functions $\tilde{F}_k (\tilde{x},
\tilde{y})$. Here $\tilde{x}$ and $\tilde{y}$ are dimensionless coordinates, and index
$k$ is in the range $k = 1, \cdots, k_{\rm max}$, where $k_{\rm max}$ is the total number
of modes. Then an arbitrary function $f(x,y,r,t)$ can be written as
\begin{equation}
f(x,y,r,t) = \sum_k f_k (r,t) \tilde{F}_k (\tilde{x},\tilde{y}) ,  \label{eq:fxy}
\end{equation}
where $f_k (r,t)$ is the amplitude of the mode with index $k$. The basis functions
depend on the dimensionless perpendicular coordinates $\tilde x \equiv x / R(r)$ and
$\tilde y \equiv y / R(r)$, where $R(r)$ is the radius of the cross-section of the flux
tube. In paper~I we assumed a circular cross-section, $\tilde{x}^2 + \tilde{y}^2 \le 1$,
but in the present work we follow \citet[][]{Perez2013} by assuming a square cross-section,
$-1 \le \tilde{x} \le +1$ and $-1 \le \tilde{y} \le +1$, and we use periodic boundary
conditions on this square domain. Then the basis functions are products of $\tilde{x}$-
and $\tilde{y}$-dependent parts, each of which are sine or cosine functions with periods
$\Delta \tilde{x} = \Delta \tilde{y} = 2$. In this case equation (\ref{eq:fxy}) is
essentially the Fourier Transform written in a compact form. The width of the computational
domain in dimensional units is $\Delta x = \Delta y = 2 R(r)$, which increases with radial
distance $r$ from Sun center. The basis functions have well-defined dimensionless
perpendicular wavenumbers $a_{x,k} = \pi n_{x,k}$ and $a_{y,k} = \pi n_{y,k}$, where
$n_{x,k}$ and $n_{y,k}$ are integers. The total dimensionless wavenumber is $a_k \equiv
\sqrt{a_{x,k}^2 + a_{y,k}^2}$, and the actual wavenumber in physical units is $k_\perp
= a_k / R(r)$. Inserting equation (\ref{eq:fxy}) into equation (\ref{eq:bracket1}),
we find for the mode amplitudes of the function $b(x,y,r,t)$:
\begin{equation}
b_k (r,t) = \frac{1}{R^2 (r)} \sum_j \sum_i M_{kji} f_j (r,t) g_i (r,t) , 
\label{eq:bracket2}
\end{equation}
where $f_j (r,t)$ and $g_i (r,t)$ are the mode amplitudes of the arbitrary functions
$f(x,y,r,t)$ and $g(x,y,r,t)$, and $M_{kji}$ is a sparse, dimensionless matrix describing
the nonlinear coupling between certain mode triples $(i,j,k)$. For the present case of
a square cross-section:
\begin{equation}
M_{kji} = \frac{1}{4} \int_{-1}^{+1} \int_{-1}^{+1} \tilde{F}_k (\tilde{x},\tilde{y})
\left( \frac{\partial \tilde{F}_j} {\partial \tilde{x}} \frac{\partial \tilde{F}_i}
{\partial \tilde{y}} - \frac{\partial \tilde{F}_j} {\partial \tilde{y}}
\frac{\partial \tilde{F}_i} {\partial \tilde{x}} \right) ~ d \tilde{x} ~ d \tilde{y} .
\label{eq:Mkji}
\end{equation}
In general the details of the $M_{kji}$ matrix depend on whether the flux tube has
a circular or square cross-section, and on the type of boundary condition used, but
the matrix is always fully antisymmetric in its indices, as was shown for the circular
case in Appendix B of \citet[][]{vanB2011}. Using equation (\ref{eq:fxy}), the RMHD
equations can be written as
\begin{eqnarray}
\frac{\partial \omega_{\pm,k}} {\partial t} & = & 
     - ( u_0 \pm v_{\rm A} ) \frac{\partial \omega_{\pm,k} } {\partial r} \nonumber \\
 & & + \frac{1}{2} \left( \frac{dv_{\rm A}} {dr} \pm \frac{u_0}{2 H_\rho} \right)
       \left( \omega_{+,k} - \omega_{-,k} \right) + \frac{u_0} {2 H_{\rm B}}
       \left( \omega_{+,k} + \omega_{-,k} \right)  \nonumber \\
 & & + \frac{1}{2 R^4} \sum_j \sum_i M_{kji} ( a_i^2 - a_j^2 - a_k^2 ) 
       f_{\pm,j} f_{\mp,i}  \nonumber \\
 & & - \nu_{\pm,k} ~ \omega_{\pm,k} , \label{eq:rmhd2}
\end{eqnarray}
where $\nu_{\pm,k}$ are artificial damping rates for dominant and minority waves.
The damping model will be described in more detail in section 3.

Multiplying equation (\ref{eq:rmhd2}) by $\onehalf \rho_0 f_{\pm,k}$ and summing
over modes, we obtain the wave energy equations for the dominant and minority waves:
\begin{equation}
\frac{\partial U_{\pm}}{\partial t} + B_0 \frac{\partial} {\partial r}
\left( \frac{F_{\pm}} {B_0} \right)  =  - u_0 D_{\pm}
\mp \onehalf \frac{dv_{\rm A}}{dr} U_{\rm R} - Q_{\pm} ,  \label{eq:energy}
\end{equation}
where 
\begin{eqnarray}
U_{\pm} & = & \frac{\rho_0} {4 R^2} \sum_{k} a_k^2 f_{\pm,k}^2 ,  \label{eq:Upm} \\
U_{\rm R} & = & \frac{\rho_0} {2 R^2} \sum_{k} a_k^2 f_{+,k} f_{-,k} , \label{eq:UR} \\
Q_{\pm} & = & \frac{\rho_0} {2 R^2} \sum_{k} \nu_{\pm,k} a_k^2 f_{\pm,k}^2 , 
  \label{eq:Qpm} \\
F_{\pm} & = & ( u_0 \pm v_{\rm A} ) U_{\pm} + \onehalf u_0 \left( U_{\pm}
  - \frac{U_{\rm R}}{2} \right) ,  \label{eq:Fpm} \\
D_{\pm} & = & - \frac{1}{2} \frac{\partial} {\partial r}
  \left( U_{\pm} - \frac{U_{\rm R}}{2} \right)  - \frac{U_{\rm R}}{2 H_B} ,
\label{eq:Dpm}
\end{eqnarray}
and we use mass conservation ($\rho_0 u_0 / B_0$ = constant). Here $U_{\pm} (r,t)$ are
the wave energy densities, $U_{\rm R} (r,t)$ is the ``residual" energy density
\citep[][]{Grappin1982, Grappin1983},
$Q_{\pm} (r,t)$ are the wave dissipation rates, $F_{\pm} (r,t)$ are the energy fluxes,
and $D_{\pm} (r,t)$ are the contributions to the wave pressure force. The total wave
energy densiy is given by $U_{\rm tot} = U_{+} + U_{-}$, and the contributions from
magnetic and kinetic energy are given by $U_{\rm mag} = ( U_{\rm tot} - U_{\rm R} )/2$
and $U_{\rm kin} = ( U_{\rm tot} + U_{\rm R} )/2$. Similarly, the total dissipation rate
$Q_{\rm tot} = Q_{+} + Q_{-}$, the total energy flux $F_{\rm tot} = F_{+} + F_{-}$,
and the total wave pressure force $D_{\rm wp} = D_{+} + D_{-}$. Note that the nonlinear
terms in the RMHD equations drop out in the energy equations (\ref{eq:energy})
(also see Appendix C of paper~I). The terms $u_0 D_{\pm}$ in the energy equations
represent the work done by the wave pressure forces on the background flow.

We now derive an expression for the energy cascade rates. For a given value $a_\perp$
of the dimensionless perpendicular wavenumber, the basis functions can be split into
two sets, a low-wavenumber set, $L \equiv \{ k ~ | ~ a_k < a_\perp \}$, and a
high-wavenumber set, $H = \{ k ~ | ~ a_k > a_\perp \}$. All wave-related quantities
have contributions from both low- and high-wavenumbers sets. For example, the wave
energy densities can be written as $U_{\pm} = U_{L,\pm} + U_{H,\pm}$, where the
subscripts $L$ and $H$ refer to the two subsets:
\begin{eqnarray}
U_{L,\pm} & = & \frac{\rho_0} {4 R^2} \sum_{k \in L} a_k^2 f_{\pm,k}^2 , \\
U_{H,\pm} & = & \frac{\rho_0} {4 R^2} \sum_{k \in H} a_k^2 f_{\pm,k}^2 .
\end{eqnarray}
Similar expressions can be written for the other quantities listed in equations
(\ref{eq:UR}) through (\ref{eq:Dpm}). Taking the time derivatives of $U_{L,\pm}$
and $U_{H,\pm}$, and using equations (\ref{eq:rmhd2}), we can derive separate energy
equations for the low- and high-wavenumber sets:
\begin{eqnarray}
\frac{\partial U_{L,\pm}}{\partial t} + B_0 \frac{\partial} {\partial r}
\left( \frac{F_{L,\pm}} {B_0} \right) & = & - u_0 D_{L,\pm}
\mp \onehalf \frac{dv_{\rm A}}{dr} U_{L,c} - Q_{L,\pm} - \epsilon_{\pm} ,
\label{eq:low} \\
\frac{\partial U_{H,\pm}}{\partial t} + B_0 \frac{\partial} {\partial r}
\left( \frac{F_{H,\pm}} {B_0} \right) & = & - u_0 D_{H,\pm}
\mp \onehalf \frac{dv_{\rm A}}{dr} U_{H,c} - Q_{H,\pm} + \epsilon_{\pm} .
\label{eq:high}
\end{eqnarray}
Here $\epsilon_{\pm}$ are the rates at which energy is transfered from set $L$ to
set $H$ by nonlinear coupling:
\begin{equation}
\epsilon_{\pm} (a_\perp,r,t) = \frac{\rho_0}{4 R^4} \sum_{k \in H} \sum_{j \in L}
\sum_i M_{kji} ( a_i^2 - a_j^2 - a_k^2 ) f_{\pm,k} f_{\pm,j} f_{\mp,i} ,
\label{eq:epsilon}
\end{equation}
where the sum over $k$ is restricted to set $H$, the sum over $j$ can be restricted
to the set $L$ (because the contributions from $j \in H$ cancel each other), and the
sum over $i$ includes all modes. These rates are functions of dimensionless
perpendicular wavenumber $a_\perp$, position $r$ along the flux tube, and time $t$.
The time-averaged cascade rates are given by
\begin{equation}
\epsilon_{\pm} (a_\perp,r) = \frac{\rho_0}{4 R^4} \sum_{k \in H} \sum_{j \in L}
\sum_i M_{kji} ( a_i^2 - a_j^2 - a_k^2 ) < f_{\pm,k} f_{\pm,j} f_{\mp,i} > ,
\label{eq:epsilon1}
\end{equation}
where $< \cdots >$ denotes a time average, and the turbulence is assumed to be in a
statistically stationary state. Note that the indices $i$, $j$ and $k$ refer to
three distinct modes ($i \ne j \ne k$). Also, the cascade rate $\epsilon_{+}$ for
the dominant waves depends linearly on the amplitude $f_{-,i}$ of the minority waves,
and conversely, the cascade rate $\epsilon_{-}$ for the minority waves depends
linearly on $f_{+,i}$. Therefore, the minority waves play an important role in the
cascade of the dominant waves, and {\it vice versa} \citep[e.g.,][]{Matthaeus1999,
Chandran2009}. 

According to equations (\ref{eq:epsilon}) and (\ref{eq:epsilon1}) the cascade rates
$\epsilon_{\pm}$ at wavenumber $a_\perp$ have contributions from all mode triples
$(i,j,k)$ that straddle the chosen wavenumber. In general there are a large number
of triples contributing to the overall cascade rate. Therefore, each term in the above
equations has only a small contribution to the overall cascade rate. This means that
the random variables $f_{\pm,k} (r,t)$, $f_{\pm,j} (r,t)$ and $f_{\mp,i} (r,t)$ are
only weakly correlated, and the triple correlation $< f_{\pm,k} f_{\pm,j} f_{\mp,i} >$
is small compared to the product of the typical values of the three mode amplitudes.
This makes it difficult to develop a statistical theory of reflection-driven wave
turbulence in an inhomogeneous atmosphere. In this paper we avoid this problem by
taking the mode amplitudes from numerical simulations. We use equation
(\ref{eq:epsilon}) to compute the time-dependent cascade rates $\epsilon_{\pm}
(a_\perp,r,t)$, and then average the results to obtain $\epsilon_{\pm} (a_\perp,r)$.
This avoids any assumptions about the statistical properties of the turbulence.

\section{WAVE-DRIVEN SOLAR WIND MODELS}

In paper~I we developed an RMHD model for wave turbulence in a thin flux tube inside
a coronal hole. The flux tube extends from the coronal base ($r_{\rm base} = 1.003$
$R_\odot$) to $r = 20$ $R_\odot$, well into the super-Alfvenic part of the wind.
In the present version of the model the flux tube has a square cross-section of size
$2 R(r)$, but we still refer to $R(r)$ as the tube radius. The method for constructing
the background atmosphere is described in Appendix D of paper~I. Briefly, the field
strength $B_0 (r)$ and temperature $T_0 (r)$ are specified analytically, and the
outflow velocity $u_0 (r)$ is computed by solving the wind equation, including the
effects of wave pressure forces on the background medium.
{\bf Three models}
for the background atmosphere will be considered:
{\bf in Models~A and B}
the plasma density $\rho_0 (r)$ and Alfv\'{e}n speed $v_{\rm A} (r)$ vary smoothly with
position along the flux tube, and
{\bf in Model~C}
there are additional density variations that simulate the effect of compressive MHD waves.
The background model includes the effects of the wave pressure force $D_{\rm wp}$ on the
outflowing plasma.
{\bf For Models~A and C}
the root-mean-square (rms) velocity of the waves at the coronal base is assumed to be
$v_{\rm rms, \odot} = 40$ $\rm km ~ s^{-1}$,
{\bf and for Model~B we use $v_{\rm rms, \odot} = 30$ $\rm km ~ s^{-1}$.}

\begin{figure}
\epsscale{1.0}
\plotone{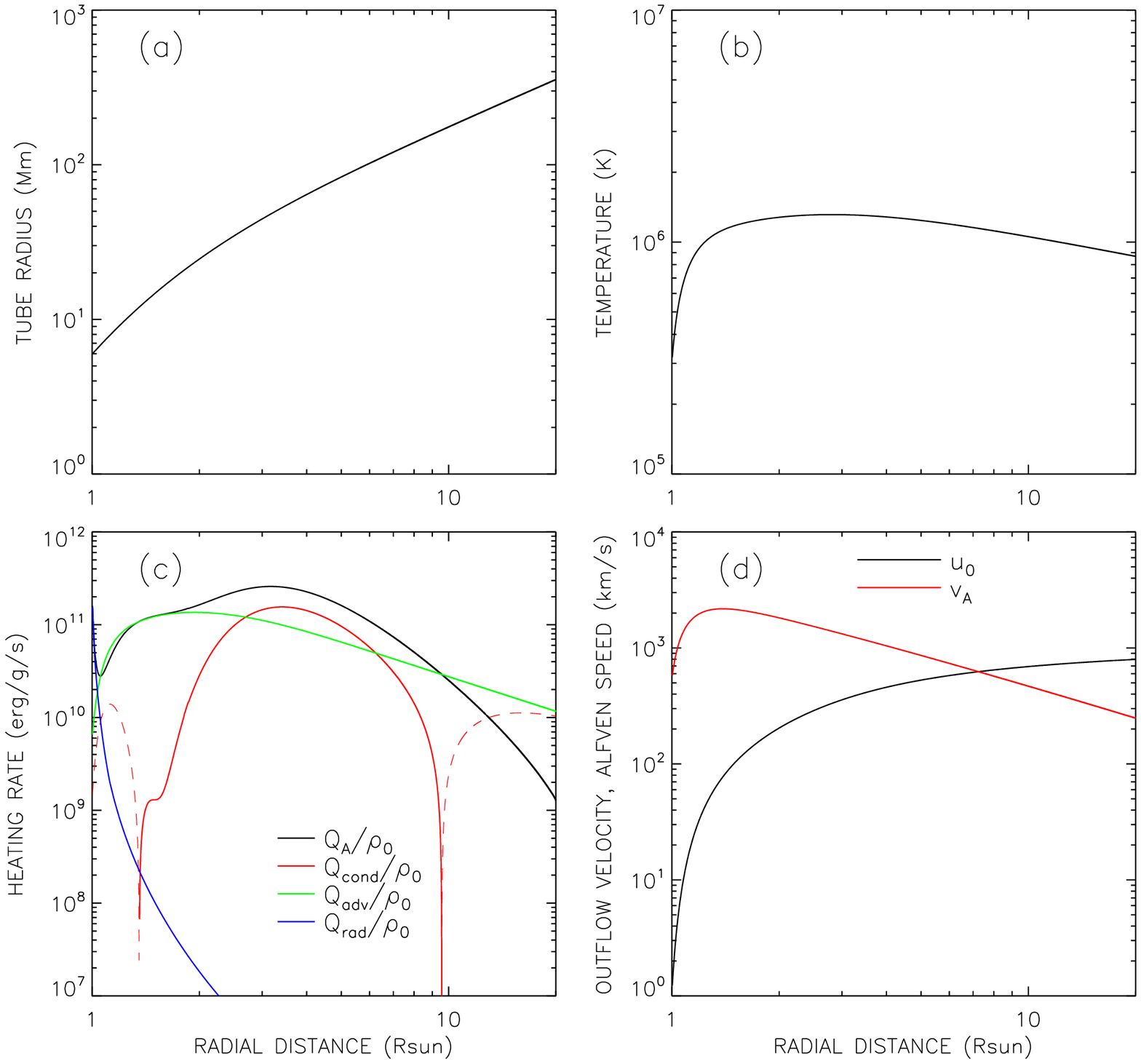}
\caption{Radial dependence of various background quantities for Model~A.
(a) Flux tube radius. (b) Temperature. (c) Plasma heating rate due to wave dissipation
(black curve), and energy-loss rates due to thermal conduction (red curve), advection
(green curve), and radiation (blue curve). (d) Outflow velocity (black curve) and
Alfv\'{e}n speed (red curve).}
\label{fig1}
\end{figure}

Figure~\ref{fig1} shows the background atmosphere for Model~A, which is nearly identical
to the model described in section 3 of paper~I. Figure~\ref{fig1}(a) shows the flux tube
radius $R(r)$, which increases from 6 Mm at the coronal base to 360 Mm at $r = 20$
$R_\odot$. Figure~\ref{fig1}(b) shows the assumed temperature $T_0 (r)$,
{\bf which is given by equation (59) of paper~I with parameter values $C_0 = 0.35$,
$C_1 = 2$, $m = 0.3$ and $k = 8$.}
Figure~\ref{fig1}(c) shows the plasma heating rate $Q_{\rm A} (r)$ necessary to maintain
this temperature. The heating is assumed to be balanced by cooling due to the expansion
of the outflowing plasma ($Q_{\rm adv}$), radiative losses ($Q_{\rm rad}$) and conductive
losses ($Q_{\rm cond}$); these contributions are shown by the colored curves in
Figure~\ref{fig1}(c).
{\bf The dashed red curves indicate regions where $Q_{\rm cond} < 0$, i.e., the convergence
of the conductive flux is heating the plasma.}
Figure~\ref{fig1}(d) shows the outflow velocity $u_0 (r)$ and the
Alfv\'{e}n speed $v_{\rm A} (r)$. Note that the Alfv\'{e}n critical point is located at
$r \approx 7.2$ $R_\odot$.

\begin{figure}
\epsscale{1.0}
\plotone{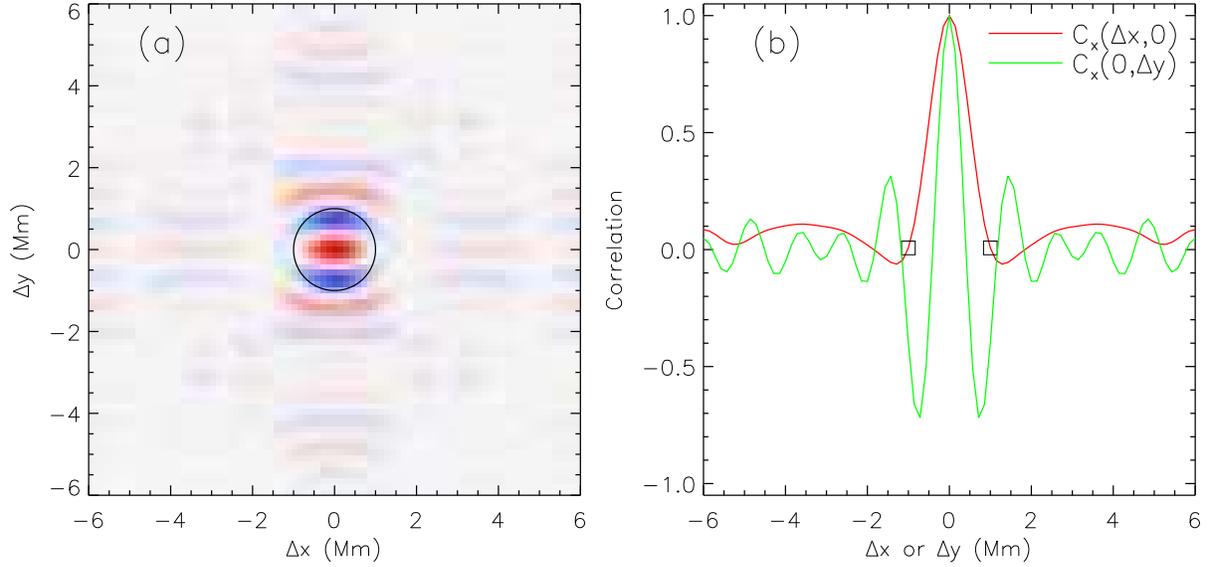}
\caption{Autocorrelation function of the footpoint velocity $v_x (x,y,t)$
at the coronal base.}
\label{fig2}
\end{figure}

On the real Sun Alfv\'{e}n- and/or kink waves may be produced by interactions of
photospheric magnetic elements with granule-scale convective flows \citep[e.g.,][]
{Spruit1982, Edwin1983, Morton2013}. Due to the density stratification of the
lower atmosphere, the waves are significantly amplified on their way to the corona.
However, the present model does not include the lower atmosphere, but starts at
the coronal base. The waves are launched by imposing random ``footpoint" motions
on the plasma and magnetic field at the coronal base. The footpoint velocity is
given by ${\bf v} = \nabla_\perp f \times \hat{\bf B}_0$, where
$f(x,y,r_{\rm base},t)$ is the velocity stream function at the coronal base.
The latter is written as a sum over basis functions:
\begin{equation}
f(x,y,r_{\rm base},t) = \sum_{k \in D} f_k (r_{\rm base},t)
\tilde{F}_k (\tilde{x},\tilde{y}) , \label{eq:fbase}
\end{equation}
where $D$ is a set of ``driver" modes with dimensionless wavenumbers in the
range $3.5 \pi < a_k < 5.5 \pi$. The amplitudes $f_k (r_{\rm base},t)$ of the driver
modes vary randomly with time $t$ in the simulation. For each mode we first create
a normally distributed random sequence $f(t)$ on a grid of times covering the entire
simulation ($t_{\rm max} =$ 30,000 s). Then the sequence is Fourier filtered using
a Gaussian function $G( \tilde{\nu} ) = \exp [-(\tau_0 \tilde{\nu} )^2]$, where
$\tilde{\nu}$ is the temporal frequency (in Hz) and $\tau_0$ is a specified parameter.
In the present work we use $\tau_0 = 120$ s, which corresponds to a correlation time
$\tau_{\rm c} = \tau_0 / \sqrt{2 \pi} \approx 48$ s; this value was chosen to be
comparable to the timescale of the solar granulation. The filtered sequences are
renormalized such that each driver mode has an equal contribution to the square of
the velocity:
\begin{equation}
\frac{a_k^2} {R_{\rm base}^2} < f_k^2 > = \frac{v_{\rm rms,\odot}^2} {N_D} ,
\label{eq:vdrv}
\end{equation}
where  $< \cdots >$ denotes a statistical average, and $N_D$ is the number of driver
modes ($N_D = 60$). The driver modes are assumed to be uncorrelated,
$< f_l f_k > = 0$ for $l \ne k$. We assume $v_{\rm rms,\odot} = 40$ $\rm km ~ s^{-1}$,
consistent with the value used in the setup of the background model. This value is also
consistent with observed spectral line widths and non-thermal velocities in coronal holes
\citep[][] {Wilhelm1998, McIntosh2008, Banerjee2009, Landi2009, Singh2011, Hahn2012,
Bemporad2012}. Note that the dynamical time of the footpoint motions is comparable to the
correlation time, $\tau_{\rm dyn} = 2 \lambda_{\perp,\odot} / v_{\rm rms,\odot} = 50$ s.
The normalized autocorrelation function for the $x$-component of velocity is given by
\begin{eqnarray}
C_x (\Delta x,\Delta y) & \equiv &
\frac{< v_x (x+\Delta x,y+\Delta y,t) v_x (x,y,t) >} {< v_x^2 >} , \nonumber \\
 & = & \frac{2}{N_D} \sum_{k \in D} \frac{a_{y,k}^2} {a_k^2} 
\cos ( a_{x,k} \Delta \tilde{x} ) \cos ( a_{y,k} \Delta \tilde{y} ) ,
\label{eq:Cx}
\end{eqnarray}
where $(\Delta x, \Delta y)$ are spatial offsets, and $(\Delta \tilde{x}, \Delta \tilde{y})$
are dimensionless values of these offsets (normalized by $R_{\rm base}$). Figure~\ref{fig2}(a)
shows this correlation function as a color-scale plot. The anisotropy of the distribution is
due to the fact that we consider here only the $x$-component of the velocity; the correlation
function for $v_y (x,y,t)$ would be rotated by 90 degrees. Figure~\ref{fig2}(b)
shows two cross-sections of the correlation function, $C_x (\Delta x,0)$ (red curve) and
$C_x (0, \Delta y)$ (green curve). The circle in Figure~\ref{fig2}(a) indicates the region
where the strongest (positive or negative) correlations are found. We use the radius of this
circle as our definition of the autocorrelation length $\lambda_{\perp,\odot}$ of the footpoint
motions. Note that $\lambda_{\perp,\odot} = 1$ Mm, significantly smaller than the domain
half-width, $R_{\rm base} = 6$ Mm.

The present model neglects all details of the collisionless processes by which the
waves are dissipated at small spatial scales. The ``dissipation range" of the turbulence
is defined as the region in wavenumber space where the simulated waves are dissipated.
In the present model the dissipation range is given by $a_k > (2/3) a_{\rm max}$,
where $a_{\rm max}$ is the maximum dimensionless wavenumber in the model.
In the region below the dissipation range we set $\nu_{\pm,k} = 0$, so that these waves
can propagate over long distance without significant dissipation. Inside the dissipation
range we use $\nu_{\pm,k} = \nu_{\pm}$, the same for all modes. The damping rate
$\nu_{\pm}$ is twice the nonlinear cascade rate for those waves:
\begin{equation}
\nu_{\pm} = 2 k_{\rm d} Z_{\mp,\rm d} ,  \label{eq:nuk}
\end{equation}
where $k_{\rm d} = (2/3) a_{\rm max} /R$ is the wavenumber at the start of the
dissipation range, and $Z_{\mp,\rm d}$ is the Elsasser variable just below this range.
The latter is given by
\begin{equation}
Z_{\mp,\rm d} \equiv \sqrt{ \sum_i (a_i /R)^2 f_{\mp,i}^2 } ,  \label{eq:zzmp}
\end{equation}
where the sum is taken over all modes with wavenumbers in the range $(1/2) a_{\rm max}
< a_i < (2/3) a_{\rm max}$. Note that the damping rate $\nu_{+}$ for the dominant
waves depends on the Elsasser variable $Z_{-,\rm d}$ of the minority waves, and
{\it vice versa}, so the damping rates satisfy $\nu_{+} \ll \nu_{-}$. This differs
from the approach used in paper~I where we assumed $\nu_{+} = \nu_{-}$.
The present method has the advantage that the waves are dissipated at a rate
comparable to the rate at which energy is injected into the dissipation range
by cascade. We find that this approach produces wave energy spectra that are only
weakly affected by the ``bottleneck" effect \citep[e.g.][]{Beresnyak2009b}.

The numerical methods for solving the RMHD equations (\ref{eq:rmhd2}) are mostly
described in Appendix B of paper~I, but there are some important differences.
In paper~I we assumed a circular cross-section, and the nonlinear terms were
evaluated by summing over the nonzero elements of $M_{kji}$. As already mentioned,
we now use a square domain and periodic boundary conditions. We use a set of modes
with $n_{x,k}$ and $n_{y,k}$ in the range 0 to 21, so the maximum dimensionless
wavenumber $a_{\rm max} = 21 \pi = 65.97$, which is higher than the value
$a_{\rm max} = 30$ used in paper~I.
In the present case the total number of modes $k_{\rm max} = 1848$, and
there are about 7 million matrix elements with $M_{kji} \ne 0$. Instead of summing
over mode triples, we evaluate the nonlinear terms using the Fast Fourier Transform
(FFT) method \citep[also see][]{Perez2013}. In essence, the arrays $f_{\pm,k}$ are
transformed into functions in real space $(\tilde{x},\tilde{y})$, and the brackets
are then computed as described in equation (\ref{eq:bracket1}). We verified that
the FFT method for computing the brackets produces exactly the same result as
explicitly summing over mode triples. The half-width $R_{\rm base}$ of the
computational domain at the coronal base is 6 Mm, significantly larger than the
correlation length  of the footpoint motions, $\lambda_{\perp,\odot} = 1$ Mm.
The RMHD equations are still integrated with a time step $\Delta t_0 = 1$ s.
The code is parallelized using OPENMP.

\section{MODELS WITH A SMOOTH BACKGROUND ATMOSPHERE}

\begin{figure}
\epsscale{1.0}
\plotone{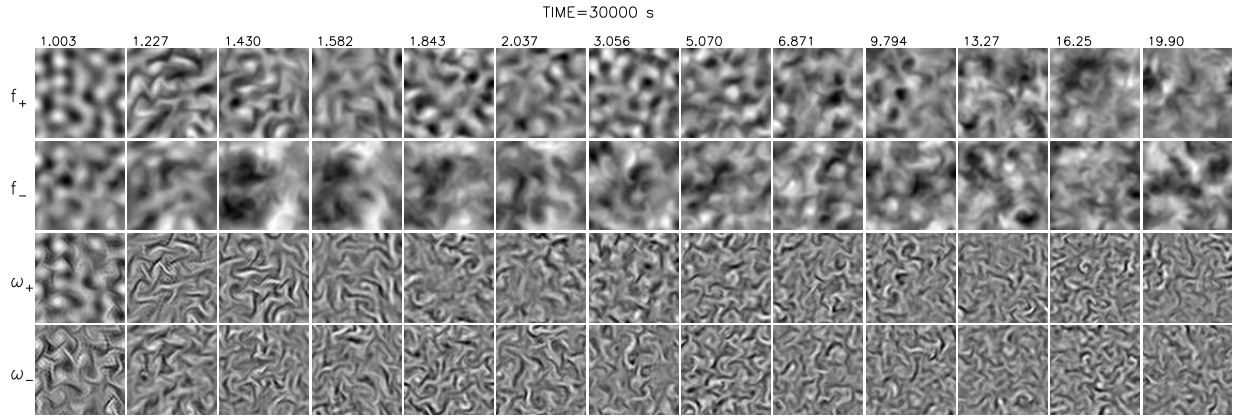}
\caption{Velocity patterns of the Alfv\'{e}n waves in cross-sections of the
flux tube. Top rows: velocity stream functions $f_{\pm} (x,y)$ of dominant
$(+)$ and minority $(-)$ waves. Bottom rows: parallel component of vorticity,
$\omega_{\pm} (x,t)$ dominant and minority waves. The different columns correspond
to different heights along flux tube. Each panel shows the normalized distribution
of the relevant quantity.}
\label{fig3}
\end{figure}

In this section we first describe RMHD simulations for Model~A, which has a smooth
background atmosphere (see Figure~\ref{fig1}). The Alfv\'{e}n waves launched at the
coronal base produce reflection-driven turbulence at larger heights.
The outward-propagating waves first reach the outer boundary of the model
($r = 20$ $R_\odot$) after about 10,859 s, and we simulate the turbulence for a period
of 30,000 s. Figure~\ref{fig3} shows wave velocity patterns in cross sections of the
flux tube at the end of the simulation. The first and second rows show the velocity
stream functions $f_{\pm} (x,y)$, and the third and fourth rows show the vorticities
$\omega_{\pm} (x,y)$. The different columns correspond to different positions along
the tube and are labeled with the radial distance $r/R_\odot$. Each panel is normalized,
so Figure~\ref{fig3} does not provide any quantitative information on the amplitude of
the waves.

\begin{figure}
\epsscale{1.0}
\plotone{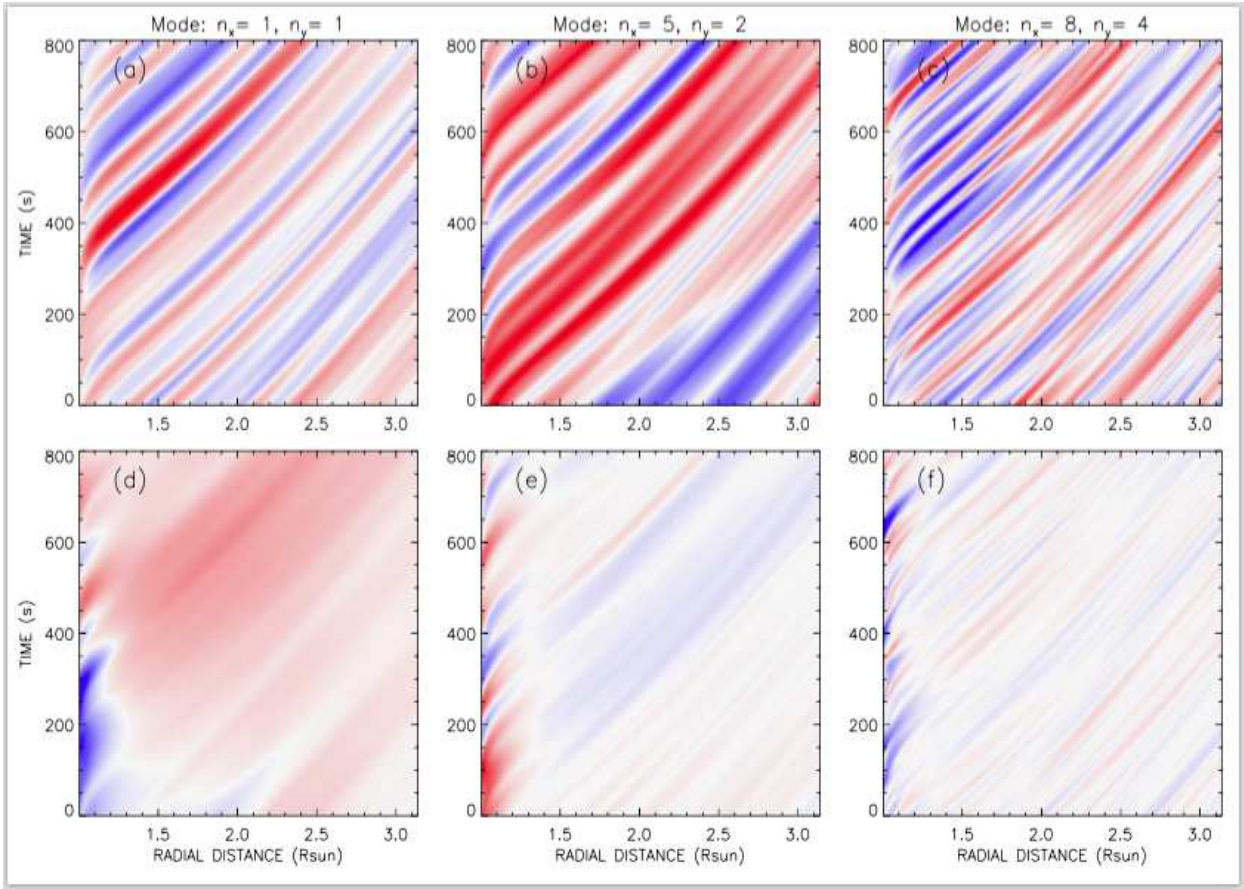}
\caption{Vorticities $\omega_{\pm,k} (r,t)$ as function of radial distance $r$ and
time $t$ for three different wave modes $k$ in Model~A, which has a smooth background
atmosphere.}
\label{fig4}
\end{figure}

\citet[][]{Velli1989} predicted that the minority waves have both an inward-propagating
``classical" component and an outward-propagating ``anomalous" component.
Figure~\ref{fig4} shows the vorticities $\omega_{\pm,k} (r,t)$ of the simulated waves
plotted as function of radial distance $r$ and time $t$ for three different wave modes $k$.
The selected modes have basis functions of the form $\tilde{F}_k (\tilde{x},\tilde{y}) =
2 \cos ( \pi n_{x,k} \tilde{x} ) \cos ( \pi n_{y,k} \tilde{y} )$, and the values of
$n_{x,k}$ and $n_{y,k}$ are given at the top of each column of Figure~\ref{fig4}.
The upper panels show the dominant waves $\omega_{+,k}$, and the lower panels show the
corresponding minority waves $\omega_{-,k}$. Note that the velocity patterns in the
lower panels have the same positive slopes as the patterns in the upper panels.
Therefore, the minority waves travel radially outward with the same velocity
($u_0 + v_{\rm A}$) as the dominant waves. There is no evidence in these diagrams for
inward-propagating waves with negative slopes, and the same is true for all wave modes
in our simulation. This means that the minority waves are dominated by the ``anomalous"
component \citep[also see][and paper~I]{Perez2013}. Therefore, it is not correct to think
of the minority waves as inward-propagating waves.

\begin{figure}
\epsscale{1.0}
\plotone{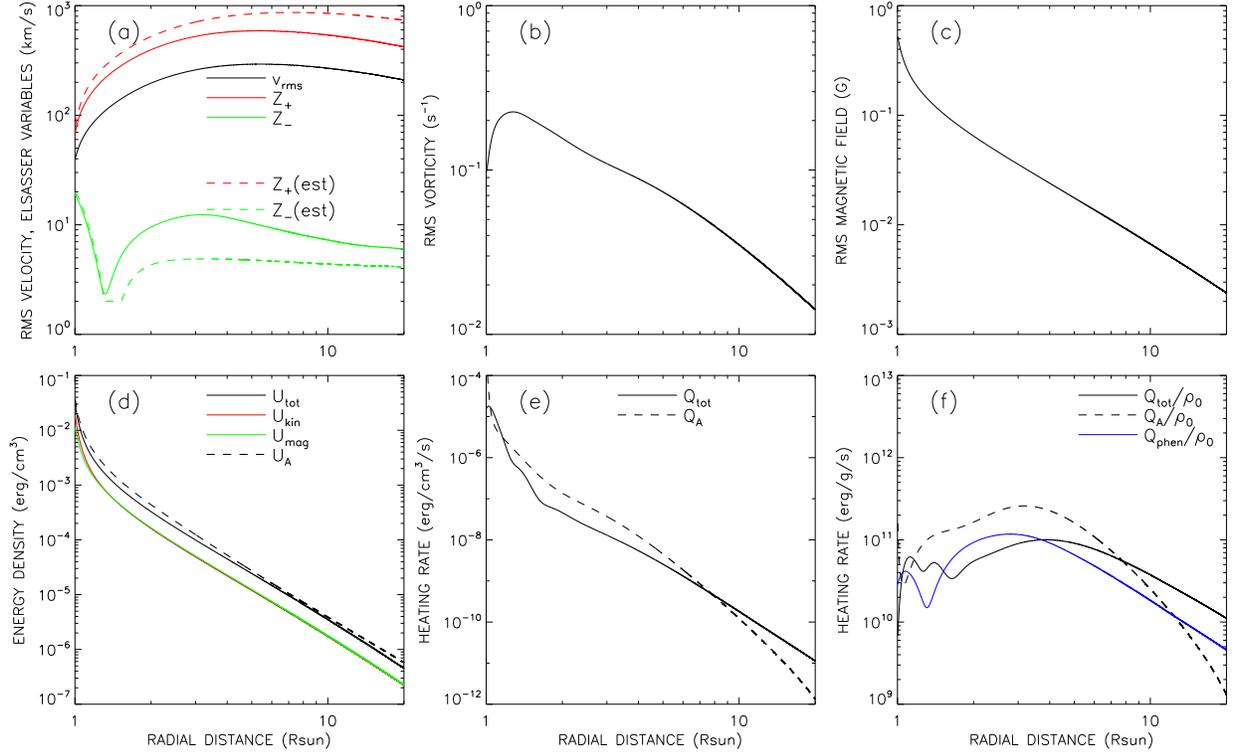}
\caption{Radial dependence of various wave-related quantities: (a) Velocity
amplitude of the waves (black curve), and Elsasser variables for dominant waves
(red curve) and minority waves (green curve). The dashed red/green curves
are estimates for the Elsasser variables. (b) Amplitude of the vorticity.
(c) Amplitude of the fluctuating component of magnetic field. (d) Wave energy
densities: total energy (black curve), kinetic energy (red curve), and magnetic
energy (green curve). Also shown is the wave energy density assumed in setup of
the background atmosphere (dashed curve). (e) Wave energy dissipation rates per
unit volume: total wave dissipation rate $Q_{\rm tot}$ (solid black curve),
and plasma heating rate $Q_{\rm A}$ assumed in setup of the background atmosphere
(dashed black curve). (f) Wave energy dissipation rates per unit mass: rate derived
from turbulence simulation (solid black curve), rate assumed in the setup of
background atmosphere (dashed curve), and rate predicted by a phenomenological
turbulence model (blue curve), equation (\ref{eq:Qphen}) with $c_{\rm d} = 0.1$.}
\label{fig5}
\end{figure}

Figure \ref{fig5} shows various wave-related quantities averaged over the
cross-section of the flux tube and over the time. Each quantity is averaged
over the time interval $t_0 (r)+300 \le t \le 30000$ (in seconds), where
$t_0 (r)$ is the time for an outward propagating wave to reach a certain height:
\begin{equation}
t_0 (r) \equiv \int_{r_{\rm base}}^{r} \frac{dr^\prime}
{u_0 (r^\prime) + v_{\rm A} (r^\prime)} .  \label{eq:t0}
\end{equation}
The black curve in Figure \ref{fig5}(a) shows the rms velocity amplitude of the
waves, $v_{\rm rms} (r)$. The {\it solid} red and green curves in Figure
\ref{fig5}(a) show the rms values of the Elsasser variables, $Z_{\pm} (r) =
\sqrt{ < | {\bf z}_\pm |^2 >}$. Note that the minority waves are much weaker
than the dominant waves; at $r > 5$ $R_\odot$ the ratio $Z_{-} / Z_{+} \approx
0.016$. The function $Z_{-} (r)$ has a sharp minimum at $r \approx 1.3$ $R_\odot$,
which is due to the fact that the Alfv\'{e}n speed has a maximum near that height,
see Figure~\ref{fig1}(d). Figure \ref{fig5}(b) shows the rms vorticity of the waves,
which is dominated by waves with high perpendicular wavenumbers and therefore more
sensitive to the spatial resolution of the model. Figure \ref{fig5}(c) shows the
rms value of the magnetic fluctuations.

The numerical results for the Elsasser variables can be compared with predictions
from a turbulence model that uses a simple phenomenology for the cascade and
dissipation of waves \citep[][]{Chandran2009a}. This analytical model gives
the following estimates \citep[also see][]{Chandran2011}:
\begin{eqnarray}
Z_{+,\rm est} & = & \frac{2 v_{\rm rms,\odot}} {1 + M_{\rm A}}
\left( \frac{\rho_0}{\rho_{0,\rm base}} \right)^{-1/4} ,  \label{eq:Zp_est} \\
Z_{-,\rm est} & = & ( 1 + M_{\rm A} ) \lambda_\perp \left| \frac{d v_{\rm A}}{dr}
\right| , 
\label{eq:Zm_est}
\end{eqnarray}
where $v_{\rm rms,\odot}$ is the velocity amplitude at the coronal base,
$\rho_{0,\rm base}$ is the base density, $M_{\rm A} (r) \equiv u_0 / v_{\rm A}$
is the Alfv\'{e}n Mach number, and $\lambda_\perp (r)$ is the perpendicular
correlation length of the turbulence. The latter is estimated by extrapolation from
the coronal base:
$\lambda_\perp (r) =  \lambda_{\perp,\odot} [ B_0 (r) / B_{\rm base} ]^{-1/2}$,
where $\lambda_{\perp,\odot} = 1$ Mm is the autocorrelation length of the footpoint
velocity at the base, and $B_{\rm base} = 10$ G is the field strength at the base.
We further impose a minimum value on the Elsasser variable for the minority waves:
$Z_{-,\rm est} > 2$ $\rm km ~ s^{-1}$. The quantities $Z_{+,\rm est}$ and
$Z_{-,\rm est}$ are plotted in Figure \ref{fig5}(a) as {\it dashed} red and green
curves. Note that these estimates are accurate to about a factor of 2.
Therefore, the model by \citet[][]{Chandran2009a} indeed provides an approximate
description of the Elsasser variables in the acceleration region of the wind.

Figure \ref{fig5}(d) shows the total energy density $U_{\rm tot}$ of the simulated
waves (full black curve), together with the contributions from the kinetic energy
$U_{\rm kin}$ (red curve) and magnetic energy $U_{\rm mag}$ (green curve).
The dashed curve shows the wave energy density $U_{\rm A}$ used in the setup of the
background model. We see that $U_{\rm kin} \approx U_{\rm mag}$ and $U_{\rm tot}
\approx U_{\rm A}$, consistent with the assumptions made in the model setup
(see paper~I).

Figure \ref{fig5}(e) shows the total energy dissipation rate $Q_{\rm tot} (r)$
of the simulated turbulence (solid black curve). Unlike for the model of paper~I,
this rate is now dominated by the contribution from damping at high perpendicular
wavenumbers, and the contribution from damping at high parallel wavenumbers is
no longer significant (but still included). The dissipation rate $Q_{\rm tot}$ is
higher than that found for the smooth model in paper~I, even though the background
atmospheres are nearly identical. This indicates that the results of paper~I are
to some degree affected by a numerical artifact, namely, a ``bottleneck" effect
that flattens the power spectrum (see Figure 3(a) of paper~I) and reduces the
cascade rate for the dominant waves. The dashed black curve in Figure~\ref{fig5}(e)
shows the plasma heating rate $Q_{\rm A} (r)$ used in the model setup. Note that
$Q_{\rm tot} < Q_{\rm A}$ over a significant height range in the model. Therefore,
the wave dissipation rate is still smaller than the plasma heating rate needed to
sustain the background atmosphere, and the model is not in thermal equilibrium. 
Figure~\ref{fig5}(f) shows the same wave dissipation and plasma heating rates per
unit mass. We also compare our results with predictions from a ``phenomenological"
turbulence model \citep[e.g.,][]{Zhou1990, Hossain1995, Matthaeus1999, Dmitruk2001},
which predicts that the dissipation rate $Q_{\rm phen}$ is given by
\begin{equation}
Q_{\rm phen} = c_{\rm d} \rho_0 \frac{Z_{+}^2 Z_{-} + Z_{-}^2 Z_{+}}
{4 \lambda_\perp} .  \label{eq:Qphen}
\end{equation}
Here $c_{\rm d}$ is a dimensionless factor of order unity. The blue curve in
Figure~\ref{fig5}(f) shows the quantity $Q_{\rm phen}/ \rho_0$ as function of radial
distance for $c_{\rm d} = 0.1$. This value was chosen to obtain a crude fit to the
actual dissipation rate $Q_{\rm tot} (r)$ predicted by the RMHD simulation. Without
this correction factor the above expression would significantly overestimate the
dissipation rate.

\begin{figure}
\epsscale{1.0}
\plotone{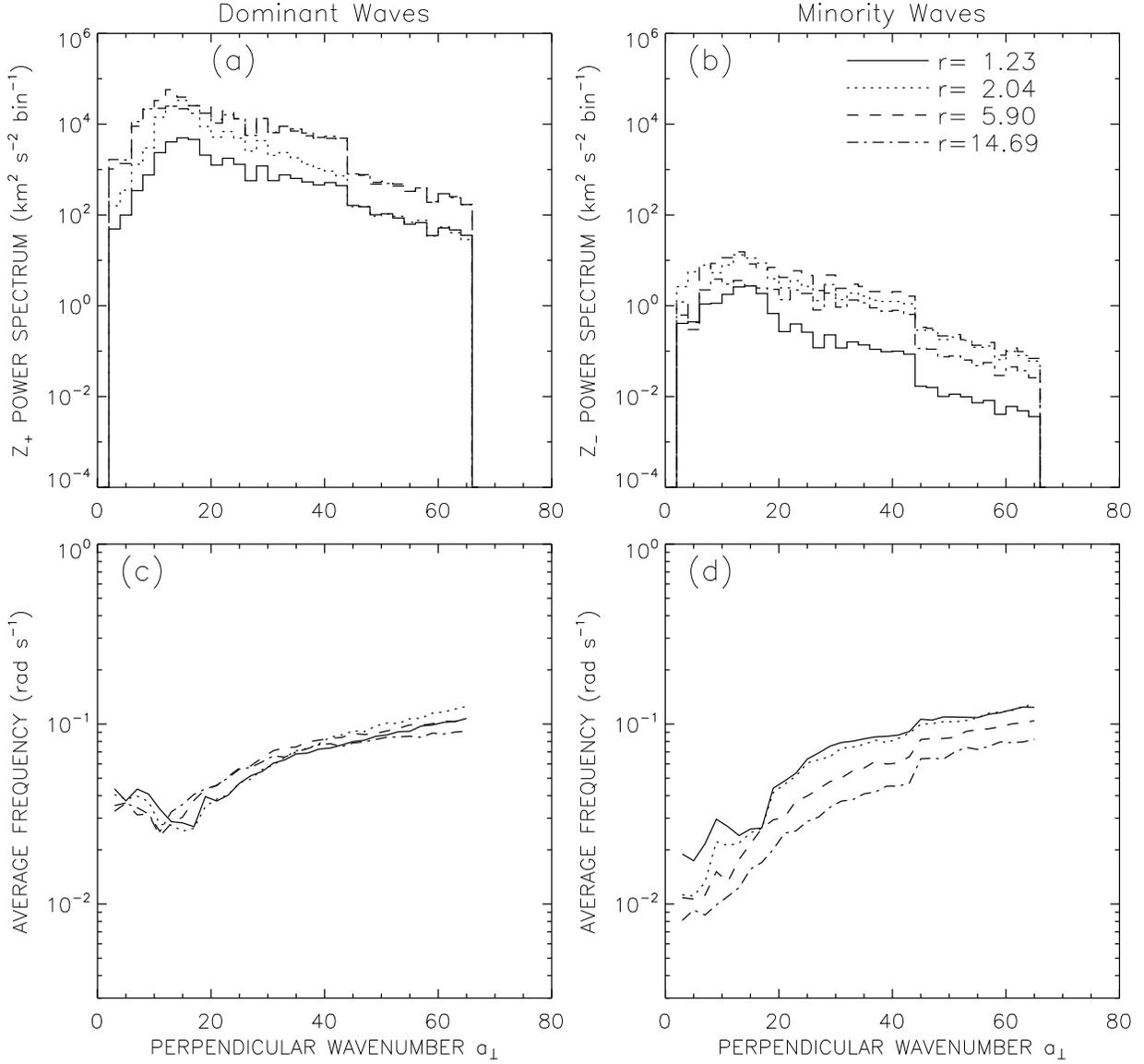}
\caption{Spatial power spectra and wave frequencies as function of
dimensionless perpendicular wavenumber $a_\perp$ for four different heights
in the model. (a) Power spectra for the Elsasser variable of the dominant waves.
The sharp drop at $a_\perp = 44$ is due to the onset of $\nu_{+,k}$-damping
at that wavenumber. (b) Power spectra for the Elsasser variable of the minority
waves. (c) Average wave frequencies for dominant waves. (d) Average wave
frequencies for minority waves. The different curves correspond to different
heights as indicated in panel (b).}
\label{fig6}
\end{figure}

Figures \ref{fig6}(a) and \ref{fig6}(b) show power spectra for the Elsasser
variables as function of dimensionless perpendicular wavenumber $a_\perp$ for
four different heights in the model. For each height we compute the wave power
in individual modes with wavenumbers $a_k$, and then collect the results into bins
in wavenumber space with $\Delta a_\perp = 2$ \citep[for details see][]{vanB2011}.
These results are derived from the last 800 s of the simulation.
Figure \ref{fig6}(a) shows the power spectra for the dominant waves.
The waves are injected at $a_\perp \sim 15$, which corresponds to the correlation
length $\lambda_{\perp,\odot}$ of the footpoint motions.
Figure \ref{fig6}(b) shows similar spectra for the minority waves.
In the present work both spectra have approximately the same slopes,
which are similar to those found in high-resolution turbulence simulations
\citep[e.g.,][]{Beresnyak2008,Beresnyak2009a, Beresnyak2009b, Perez2009,
Perez2012, Perez2013}.
We also compute {\it temporal} power spectra of dominant and minority waves,
and derive the average wave frequency $\tilde{\omega}_{\pm}$ as function of
dimensionless perpendicular wavenumber $a_\perp$. The results are shown
in Figures \ref{fig6}(c) and \ref{fig6}(d) for four different heights in
the model.

\begin{figure}
\epsscale{1.0}
\plotone{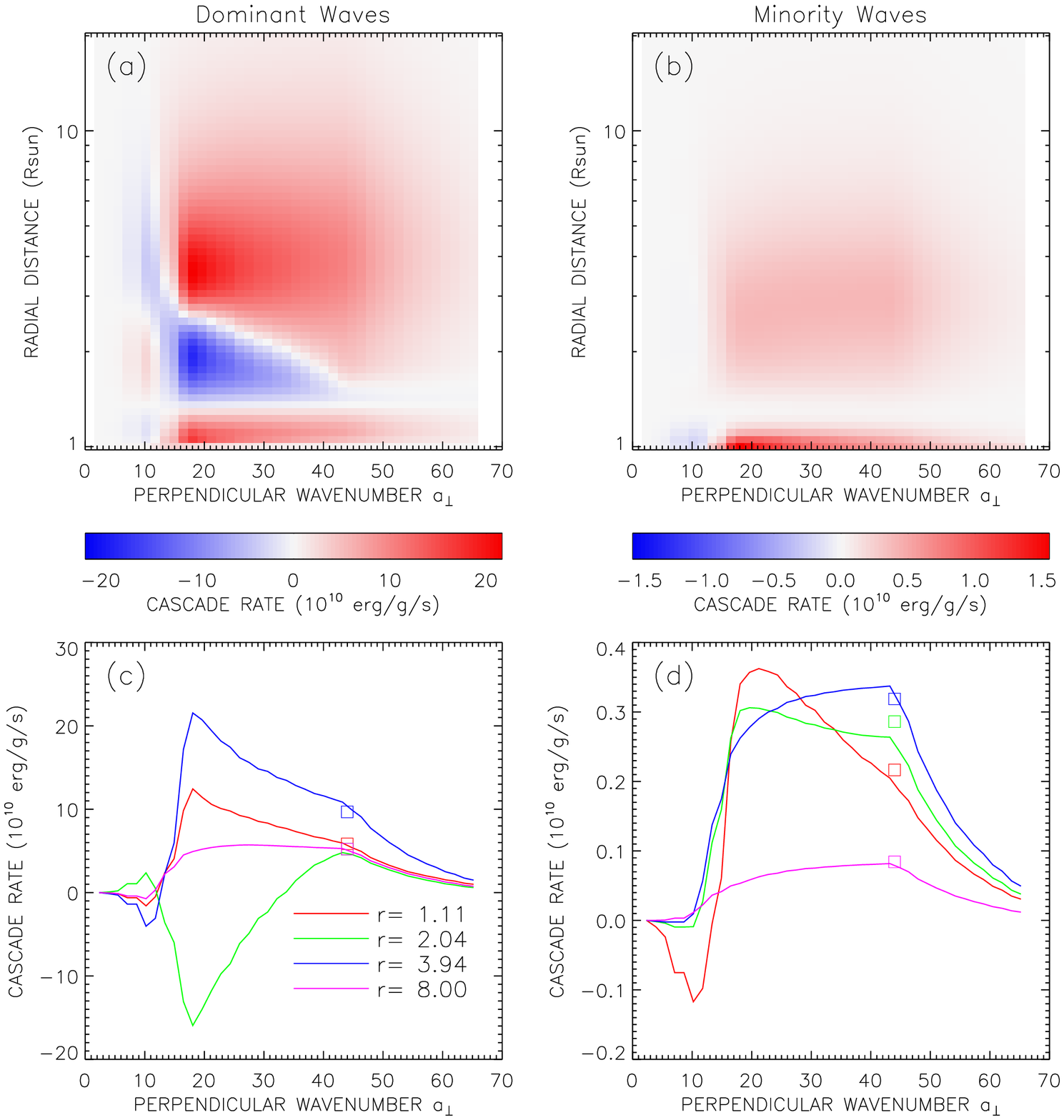}
\caption{Energy cascade rates in the simulated turbulence for Model~A.
(a) Cascade rate per unit mass for the dominant waves, $\epsilon_{+}/\rho_0$,
as function of dimensionless perpendicular wavenumber $a_\perp$ and radial
distance $r$ from Sun center. Direct and inverse cascades are indicated by red
and blue colors, respectively (see color bar).
(b) Cascade rate per unit mass for the minority waves, $\epsilon_{-}/\rho_0$,
with separate color bar.
(c) Dominant cascade rates as function of $a_\perp$ for four different radial
distances, as indicated by the legend.
(d) Minority cascade rates at the same heights.
}
\label{fig7}
\end{figure}

Figure~\ref{fig7} shows the time-averaged energy cascade rates per unit mass,
$\epsilon_{\pm}/\rho_0$. These rates are functions of dimensionless perpendicular
wavenumber $a_\perp$ and radial distance $r$. Figures~\ref{fig7}(a) and
\ref{fig7}(b) show color-scale plots for the dominant and minority waves,
respectively. Note that each plot has its own color bar, and that the cascade
rates for the dominant waves are much larger than those for the minority waves.
Red and blue colors indicates direct ($\epsilon_{\pm} > 0$) and inverse
($\epsilon_{\pm} < 0$) cascades, respectively. Figure~\ref{fig7}(a) shows that
at low heights the dominant waves have a direct cascade, but $\epsilon_{+}$ changes
sign at $r = 1.4$ $R_\odot$ for wavenumbers in the range $15 < a_\perp < 44$.
The lower boundary of this range ($a_\perp = 15$) is approximately where energy
is injected into the turbulence by the driver waves, and the upper boundary
($a_\perp = 44$) is where the waves start to be dissipated. The inverse cascade
continues up to $r = 2.5$ $R_\odot$, but in a narrowing wavenumber range.
These negative cascade rates reduce the amount of energy that can cascade into
the dissipation range ($a_\perp > 44$), and therefore affect the overall wave
dissipation rate between 1.4 and 2.5 $R_\odot$. There is also a further extension
of the region of inverse cascade to larger heights ($r > 2.5$ $R_\odot$) and low
perpendicular wavenumbers ($a_\perp < 15$), but this feature is relatively weak
and does not seem to have a strong effect on the dissipation rates. In contrast,
Figure~\ref{fig7}(b) shows that the minority waves have a direct cascade at all
heights.

Figures~\ref{fig7}(c) and \ref{fig7}(d) show plots of the cascade rates,
$\epsilon_{\pm}/\rho_0$, as function of perpendicular wavenumber for four
different heights.
These heights were chosen to represent the region of direct cascade near the
coronal base ($r = 1.11$ $R_\odot$, red curves), the region of inverse cascade at
intermediate heights ($r = 2.04$ $R_\odot$, green curves), and the direct cascades
at large heights ($r = 3.94$ $R_\odot$, blue curves; $r = 8.0$ $R_\odot$, magenta
curves). The colored squares give the wave dissipation rates $Q_{\pm}/\rho_0$.
The dissipation rates are approximately equal to the cascade rates at the start
of the dissipation range ($a_\perp = 44$), where the squares are plotted. Hence,
the energy that cascades into the dissipation range is indeed dissipated shortly
afterward, as expected. Note that at most heights the cascade rates vary strongly
with wavenumber $a_\perp$, even in the ``inertial" range ($15 < a_\perp < 44$) where
no injection or dissipation of the energy occurs. This indicates that the transport
of wave energy is a highly non-local process: as the waves cascade, they also
propagate upward in height over a significant distance. This is due to the fact
that the nonlinear cascade time for the dominant waves is comparable to the wave
propagation time (see paper~I).

What is the cause of the inverse cascade for the dominant waves? To answer this
question we first consider the production of minority waves by ``reflection" of
dominant waves. Converting vorticities to mode amplitudes $f_{\pm,k}$ and using
the fact that the minority waves are weak ($| f_{-,k} | \ll | f_{+,k} |$),
equation (\ref{eq:rmhd2}) can be written as
\begin{eqnarray}
\frac{\partial f_{-,k}} {\partial t}
+ ( u_0 - v_{\rm A} ) \frac{\partial f_{-,k}} {\partial r} & \approx &
\onehalf ( 1 + M_{\rm A} ) \frac{dv_{\rm A}}{dr} f_{+,k}  \nonumber \\
 & & + \frac{1}{2 R^2 a_k^2} \sum_j \sum_i M_{kji} ( a_i^2 - a_j^2 - a_k^2 ) 
        f_{-,j} f_{+,i} ,  \label{eq:rmhd3}
\end{eqnarray}
where we omit the wave damping terms. Note that the ``nonlinear" term in equation
(\ref{eq:rmhd3}) is in fact linear in the amplitudes $f_{-,j} (r,t)$ of the minority
waves. Also, the production of minority waves is proportional to the Alfv\'{e}n
speed gradient $dv_{\rm A} /dr$, which is positive at heights below the peak in
Alfv\'{e}n speed ($r < 1.4$ $R_\odot$) and negative above the peak ($r > 1.4$
$R_\odot$). The dominant waves have a long cascade time and evolve only gradually
with height. However, the minority waves have a short cascade time (about 10 s at
the outer scale of the turbulence, see paper~I), and respond much more rapidly
to the changes in $dv_{\rm A} /dr$ with height. Therefore, as the waves propagate
outward through the region around the peak in Alfv\'{e}n speed, the dominant
waves $f_{+,k} (r,t)$ remain more or less unchanged, while the minority waves
$f_{-,k} (r,t)$ change sign. This reversal of the minority waves at $r \approx 1.4$
$R_\odot$ occurs for most modes, and can be seen in diagrams such as
Figure~\ref{fig4}. The reversal occurs even when the dominant and minority waves
are not well correlated with each other, as is the case at higher wavenumbers.
We now consider two heights, one just below the peak in Alfv\'{e}n speed ($r = r_1$)
and another just above it ($r = r_2$), such that the magnitudes of the gradients
are the same at the two heights: $( dv_{\rm A} /dr )_2 = - (dv_{\rm A} /dr )_1$.
The dominant and minority waves at these heights are approximately related by
\begin{eqnarray}
f_{+,k} (r_2 ,t) & \approx & f_{+,k} (r_1 ,t - \Delta t_{12}) , 
\label{eq:fp2} \\
f_{-,k} (r_2 ,t) & \approx & - c_0 ~ f_{-,k} (r_1 ,t - \Delta t_{12}) ,
\label{eq:fm2}
\end{eqnarray}
where $c_0 \approx 1$ (independent of mode index $k$), and $\Delta t_{12}$ is the
wave propagation time between the two heights. These relationships follow from
a symmetry of equation (\ref{eq:rmhd3}): the equation remains valid when the signs
of $dv_{\rm A} /dr$ and all $f_{-,k}$ are reversed, but $f_{+,k}$ is unchanged.
Inserting expressions (\ref{eq:fp2}) and (\ref{eq:fm2}) into equation
(\ref{eq:epsilon1}), we find the following relationships between the cascade rates
at the two heights:
\begin{eqnarray}
\epsilon_{+} (a_\perp,r_2) & \approx & - c_0 ~ \epsilon_{+} (a_\perp,r_1) ,
\label{eq:ep2} \\
\epsilon_{-} (a_\perp,r_2) & \approx & c_0^2 ~ \epsilon_{-} (a_\perp,r_1) .
\label{eq:em2}
\end{eqnarray}
Figure~\ref{fig7} indicates that at low heights there is a direct cascade for both
wave types, $\epsilon_{\pm} (a_\perp,r_1) > 0$. Then equations (\ref{eq:ep2}) and
(\ref{eq:em2}) predict that at $r = r_2$ there is an inverse cascade for the dominant
waves and a direct cascade for the minority waves, $\epsilon_{+} (a_\perp,r_2) < 0$
and $\epsilon_{-} (a_\perp,r_2) > 0$. The inverse cascade of the dominant waves occurs
at all wavenumbers $a_\perp$ because the reversal in sign of $f_{-,k}$ is rapidly
transmitted to larger wavenumbers by the direct cascade of the minority waves.

As the dominant waves propagate farther out they gradually adjust to the condition
$dv_{\rm A} /dr < 0$, and $\epsilon_{+}$ becomes positive again. The adjustment
of the dominant waves occurs first at large wavenumbers where the cascade times
for the dominant waves are shortest, and later also at smaller wavenumbers. Therefore,
in Figure~\ref{fig7}(a) the upper boundary of the region with inverse cascade
lies at an angle in the $(a_\perp, r)$ plane. We conclude that the inverse
cascade in Figure~\ref{fig7}(a) is linked to the change of sign of $dv_{\rm A} /dr$
at $r = 1.4$ $R_\odot$, together with the fact that the cascade time for the
dominant waves is relatively large and comparable to the wave travel time $t_0 (r)$.

\begin{figure}
\epsscale{1.0}
\plotone{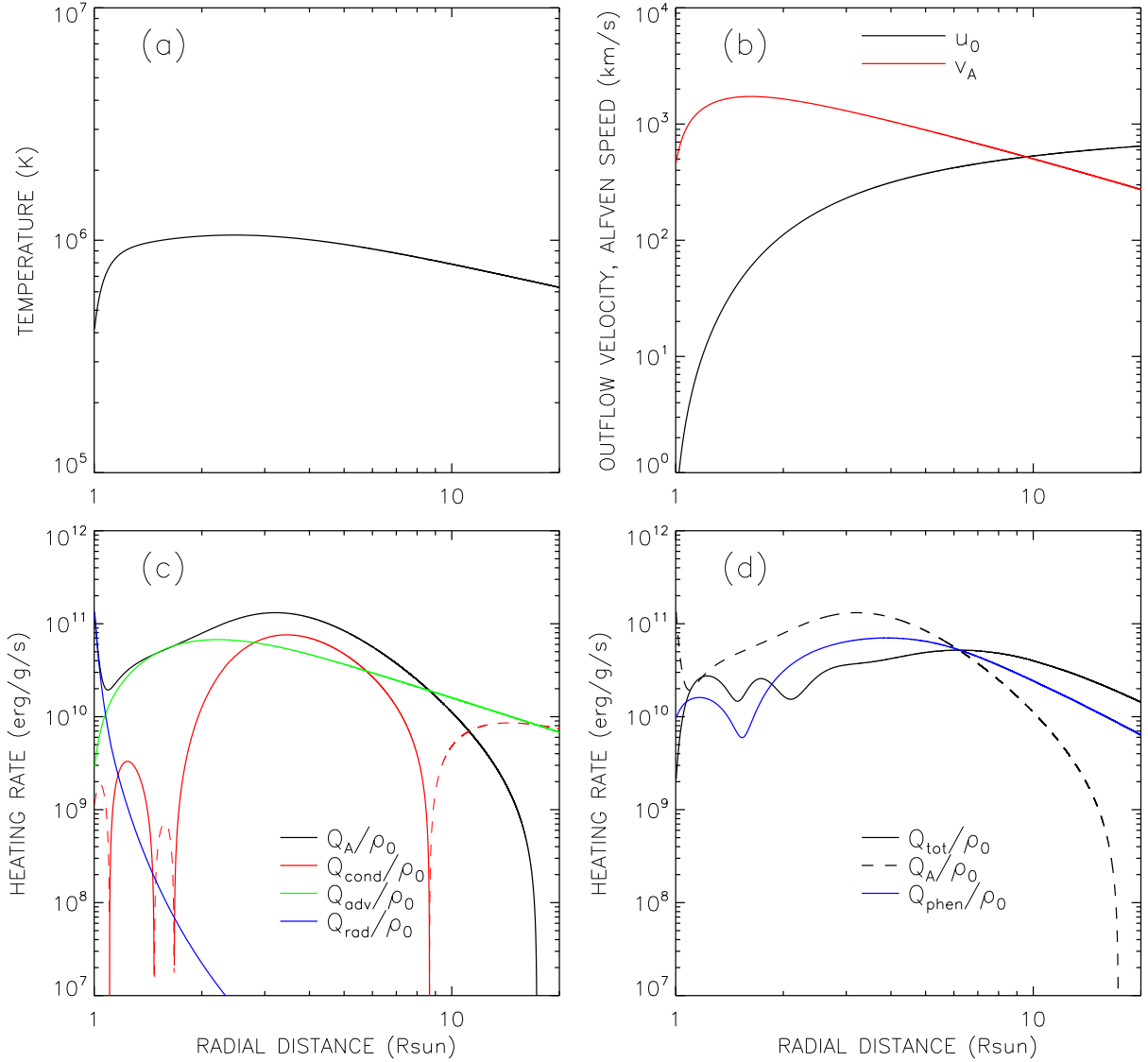}
\caption{Background quantities and simulation results for Model~B.
(a) Temperature. (b) Outflow velocity (black curve) and Alfv\'{e}n speed (red curve).
(c) Plasma heating rate required to maintain the background atmosphere (black curve),
and energy-loss rates due to thermal conduction (red curve), advection (green curve)
and radiation (blue curve). (d) Wave energy dissipation rate as derived from the RMHD
simulation (solid black curve), rate assumed in the setup of background atmosphere
(dashed curve), and rate predicted by a phenomenological turbulence model (blue curve),
equation (\ref{eq:Qphen}) with $c_{\rm d} = 0.1$.}
\label{fig8}
\end{figure}

{\bf
The background atmosphere for Model~A was chosen to be the same as that used
in paper~I, so we could directly compare our results and understand why the model of
paper~I gives such low wave dissipation rates. However, the outflow speed in this model
reaches 800 $\rm km ~ s^{-1}$ by 20 $R_\odot$ (see Figure~\ref{fig1}(d)), which is high
considering that further acceleration may occur at larger radii. Also, the Alfv\'{e}n
critical point is located at 7.2 $R_\odot$, which is low compared to other models that
rely on Alfv\'{e}n waves to heat and accelerate the fast solar wind \citep[e.g.,][]
{Cranmer2007, Verdini2010, Chandran2011}. Therefore, we now consider an alternative,
Model~B, which has a different set of model parameters. Three of the four parameters
describing the background temperature were modified ($C_0 = 0.30$, $m = 0.35$,
$k = 12$, see Appendix D in paper~I), which leads to a reduction of the peak temperature
from 1.31 MK to 1.05 MK.
The revised temperature profile is shown in Figure~\ref{fig8}(a). We also increased
the coronal base pressure from 0.1 to 0.2 $\rm dyne ~ cm^{-2}$, and decreased the
wave amplitude to $v_{\rm rms, \odot} = 30$ $\rm km ~ s^{-1}$, which reduces the
wave pressure acceleration. Figure~\ref{fig8}(b) shows the outflow velocity $u_0 (r)$
and Alfv\'{e}n speed $v_{\rm A} (r)$ resulting from these changes. In Model~B the
peak in Alfv\'{e}n speed occurs at $r \approx 1.6$ $R_\odot$, and the Alfv\'{e}n
critical point is located at $r = 9.6$ $R_\odot$, more in line with the values used
in earlier models. The plasma heating rate $Q_{\rm A} (r)$ needed to maintain the
background temperature is shown by the black curve in Figure~\ref{fig8}(c), together with
the cooling rates due to radiation (blue curve), thermal conduction (red curve) and
solar wind expansion (green curve). Comparison with Figure~\ref{fig1}(c) for Model~A
shows that the heating rate is significantly reduced. In fact, at large radii
$Q_{\rm A}$ becomes slightly negative, which is due to the near balance of conduction
heating and expansion cooling at $r = 20$ $R_\odot$ in Model~B.

We simulated the dynamics of the Alfv\'{e}n waves in Model~B for a period of 30,000 s.
The imposed footpoint motions are the same as in Model~A. The full black curve in
Figure~\ref{fig8}(d) shows the time-averaged wave dissipation rate $Q_{\rm tot} (r)$,
and the blue curve shows the rate $Q_{\rm phen} (r)$ predicted by the phenomenological
turbulence model, equation (\ref{eq:Qphen}) with $c_{\rm d} = 0.1$. Note that the
wave dissipation rate $Q_{\rm tot}$ is reduced compared to Model~A, and is below
the required heating rate $Q_{\rm A}$ (dashed curve) for $1.5 < r < 5$ $R_\odot$.
This is again due to the presence of an inverse cascade, in this case in the
height range $1.6 < r < 4$ $R_\odot$. The heating produced by the waves is insufficient
to maintain the assumed background temperature, so the model is not in thermal
equilibrium. }

\section{MODEL WITH DENSITY VARIATIONS}

In this section we describe simulation results for Model~C, which has a background
atmosphere with additional density variations along the flux tube. These variations
simulate the effect that compressive MHD waves may have on the propagation and
reflection of Alfv\'{e}n waves. The density variations $\delta \rho_0 (r)$ are
assumed to be random in position, but constant in time, consistent with our RMHD
methodology. The present model is similar to that described in section 4 of paper~I,
but the model parameters are different.
{\bf 
The magnitude of the density variations has been doubled (to $\epsilon_{\rm rms}
= 0.2$), and the correlation length of the variations has been increased by a factor 5
(to $\lambda_{\rm c} = 0.2$ $R_\odot$). Therefore, the magnitude of the variations in
Alfv\'{e}n speed gradient $d v_{\rm A} /dr$ has decreased by a factor 2.5, reducing the
wave reflection. The main reason for these changes is to bring
out more clearly the spatial variations of the cascade rates $\epsilon_{\pm}$ of the
dominant- and minority waves. In reality there are density fluctuations both along and
perpendicular to the field lines, and the latter may actually be much larger than the
former (see references in section~4 of paper~I). We believe that perpendicular density
fluctuations will have an important effect on the cascade rates, but unfortunately we
are unable to simulate this effect with our RMHD code, which assumes constant density
over the cross-section of the flux tube. To compensate, we use a high value for the
magnitude of the density fluctuations along the field lines. This approach is rather
artificial, but it is the best we can do right now. }

\begin{figure}
\epsscale{1.0}
\plotone{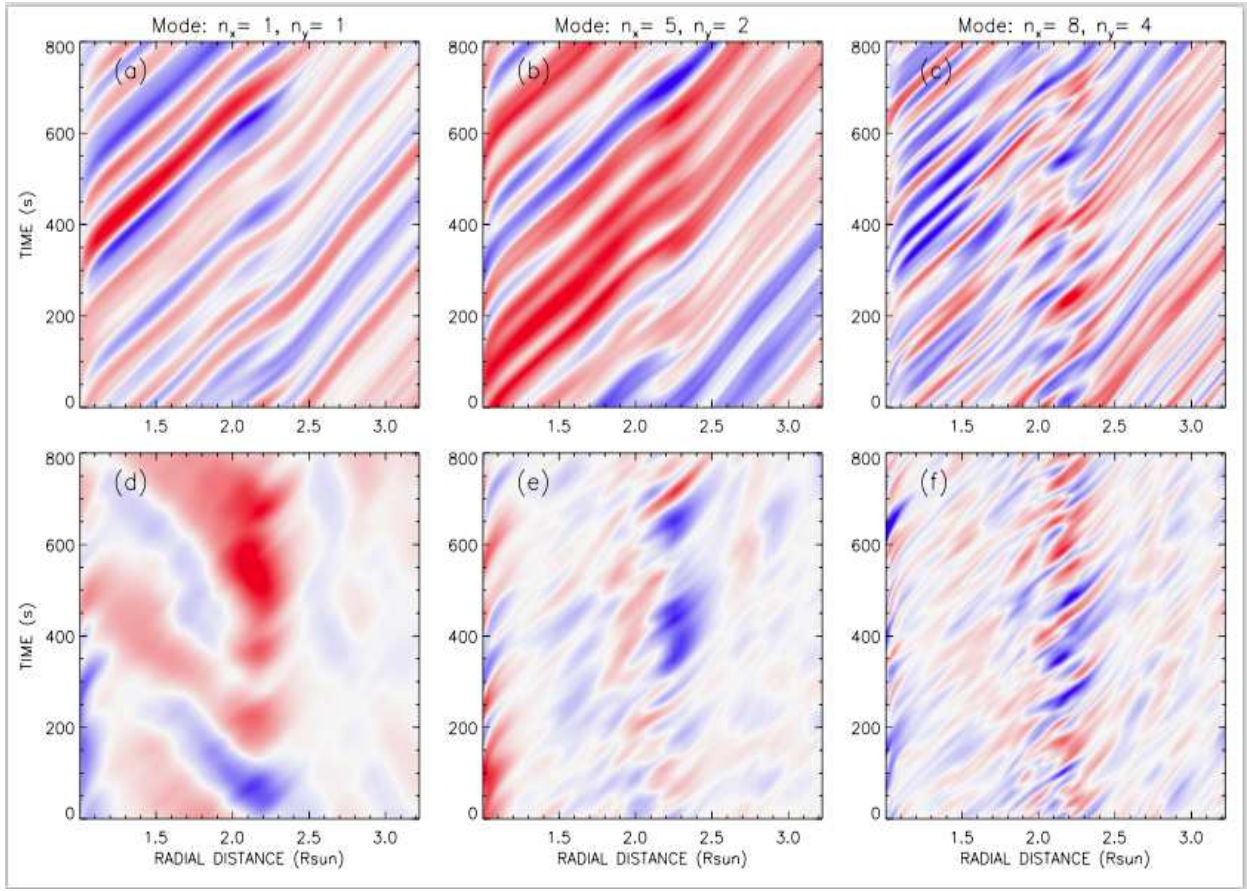}
\caption{Vorticities $\omega_{\pm,k} (r,t)$ as function of radial distance $r$ and
time $t$ for three different wave modes $k$ in Model~C with random density
variations.}
\label{fig9}
\end{figure}

In model~C the Alfv\'{e}n speed gradient $dv_{\rm A} /dr$ changes sign multiple
times with increasing $r$, which significantly enhances the wave reflection.
Figure~\ref{fig9} shows the vorticities $\omega_{\pm,k} (r,t)$ for three different
wave modes, the same modes as in Figure~\ref{fig4}. The upper panels of
Figure~\ref{fig9} show the dominant waves $\omega_{+,k}$, and the lower panels show
the corresponding minority waves $\omega_{-,k}$. The positive slopes in the upper
panels indicate that the dominant waves travel radially outward with velocity
$u_0 + v_{\rm A}$. However, the velocity patterns in Figure~\ref{fig9}(d) have negative
slope, indicating this low-wavenumber minority wave ($a_k = 4.44$) is propagating
radially inward with velocity $u_0 - v_{\rm A} < 0$; this is the ``classical" component
of the minority waves described by \citet[][]{Velli1989}. In Figures~\ref{fig9}(e) and
\ref{fig9}(f) the minority waves have mostly outward-propagating components.
All panels show a stationary (vertical) pattern, which is an artifact of our assumption
that the density variations $\delta \rho_0 (r)$ are constant in time. At high wavenumbers
the minority waves are uncorrelated with the dominant waves (compare Figures~\ref{fig9}(f)
and \ref{fig9}(c)). These patterns are quite different from those for the smooth Model~A
(see Figure~\ref{fig4}).

\begin{figure}
\epsscale{1.0}
\plotone{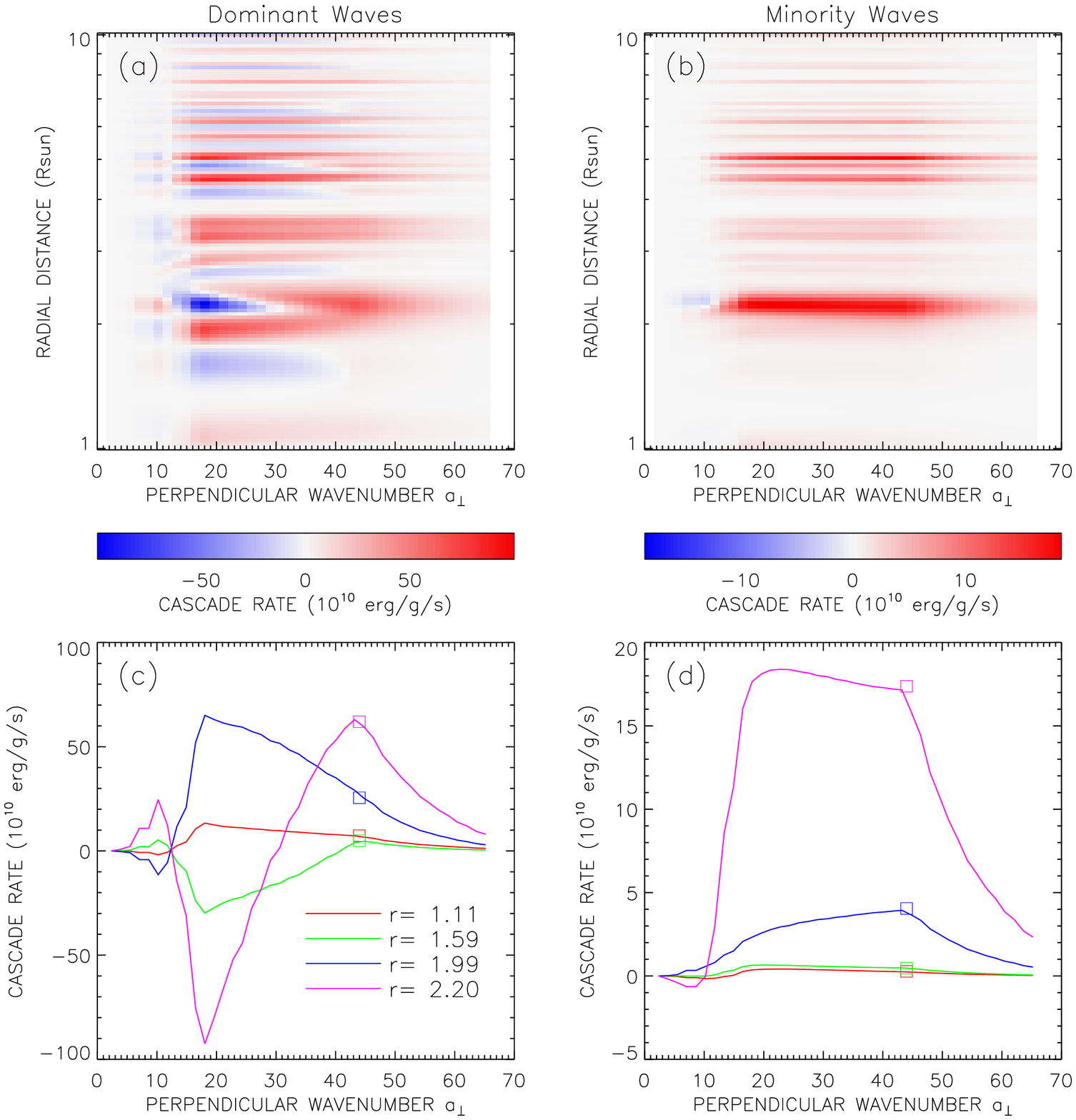}
\caption{Energy cascade rates in the simulated turbulence for Model~C.
(a) Cascade rate per unit mass for the dominant waves, $\epsilon_{+}/\rho_0$,
as function of dimensionless perpendicular wavenumber $a_\perp$ and radial
distance $r$ from Sun center. Direct and inverse cascades are indicated by red
and blue colors, respectively (see color bar).
(b) Cascade rate per unit mass for the minority waves, $\epsilon_{-}/\rho_0$,
with separate color bar.
(c) Dominant cascade rates as function of $a_\perp$ for four different radial
distances, as indicated by the legend.
(d) Minority cascade rates at the same heights.
}
\label{fig10}
\end{figure}

Figure~\ref{fig10} shows the time-averaged cascade rates $\epsilon_{\pm} (a_\perp,r)$
for the model with density variations. Figure~\ref{fig10}(a) shows the cascade rate
$\epsilon_{+} / \rho_0$ for the dominant waves. Note that with increasing height $r$ the
cascade rate changes sign multiple times, going from a direct cascade (red) to inverse
cascade (blue) over short distances. In contrast, the minority waves always
have a direct cascade, see Figure~\ref{fig10}(b). The changes in $\epsilon_{+}$ as 
function of $r$ are due to changes in $dv_{\rm A} /dr$, but the dominant waves try to
adjust to these changes, so at heights above 3 $R_\odot$ the quantities $\epsilon_{+}$
and $dv_{\rm A} /dr$ are only poorly correlated.
Comparison of the color bars in Figures~\ref{fig7} and \ref{fig10} indicate that the
magnitudes of the cascade rates in Model~C are much larger than those in Model~A.
Figures~\ref{fig10}(c) and \ref{fig10}(d) show the cascade rates $\epsilon_{\pm}/\rho_0$
as functions of wavenumber for four different heights. The colored squares indicate the
wave dissipation rates $Q_{\pm}/\rho_0$. Note that for the dominant waves the cascade
rates vary strongly with wavenumber, and the dissipation rates $Q_{+}$ are much smaller
than the peak cascade rates $| \epsilon_{+} |_{\rm max}$, which generally occur at low
wavenumber. This indicates that the rapid changes in $\epsilon_{+}$ as function of $r$
prevent the efficient cascade of wave energy to higher wavenumbers, and thereby have a
negative effect on the dissipation rate. The total dissipation rate $Q_{\rm tot} (r)$
for Model~C is slightly larger than that for Model~A, but is still insufficient to
maintain the temperature of the background atmosphere. Therefore, Model~C is also not
in thermal equilibrium.

\section{DISCUSSION AND CONCLUSIONS}

In this work we simulate the dynamics of Alfv\'{e}n waves
{\bf for three models of the fast solar wind, two with a smooth background atmosphere
(Models~A and B) and one with density fluctuations (Model~C).}
These models are improved versions of the models presented in paper~I. We compute for the
first time the energy cascade rates $\epsilon_{\pm} (a_\perp,r)$ for dominant and minority
waves, and find that at certain heights and wavenumbers the dominant waves undergo an inverse
cascade, $\epsilon_{+} < 0$. This means that the nonlinear interactions between dominant and
minority waves cause energy to be transported from smaller to larger scales, opposite to the
direction usually assumed for Alfv\'{e}n wave turbulence. Inverse cascades are predicted
to occur in two-dimensional magneto-hydrodynamic systems \citep[][]{Fyfe1976, Fyfe1977},
and in three-dimensional systems with large magnetic helicity \citep[][]{Frisch1975,
Dmitruk2007}. In the present case the inverse cascade appears to be due to a different
mechanism, namely, the change in sign of the gradient of the Alfv\'{e}n speed $dv_{\rm A} /dr$
as function of height in the model. In Model~A the inverse cascade occurs for radii between
1.4 to 2.5 $R_\odot$ and wavenumbers in the inertial range ($15 < a_\perp < 44$);
{\bf in Model~B the inverse cascade occurs between 1.6 to 4 $R_\odot$.}
In Model~C with density fluctuations the cascade rate changes sign multiple times as function
of height. In all models the cascade rate $\epsilon_{+} (a_\perp,r)$ varies significantly with
perpendicular wavenumber. This indicates that wave propagation plays an important role in the
cascade process. The cascade time scale for the dominant waves is comparable to the wave travel
time $t_0 (r)$, so in one cascade time the waves travel a distance comparable to the radial
distance $r$. The energy injected into the cascade by the driver waves at position $r$ is not
dissipated locally, but is dissipated only much later at significantly larger heights.
Therefore, the cacsade of the dominant waves is a highly non-local process.

The inverse cascade impedes the efficient transport of wave energy to large perpendicular
wavenumbers, and thereby has a negative effect on the wave dissipation rate. Therefore,
the low dissipation rates found here
{\bf (and in paper~I)}
are a real physical effect, not a numerical artifact.
In both the smooth models and the model with density fluctuations there is a significant
height range where $Q_{\rm tot} (r)$ is less than the plasma heating rate $Q_{\rm A} (r)$
needed to maintain the temperature of the background atmosphere. Hence, these models are
not self-consistent from an energy point of view. To obtain a model with higher dissipation
rates would require that the inverse cascade is somehow avoided, or at least reduced in
magnitude. At present it is unclear how to construct such a model.

{\bf 
In the smooth models the total dissipation rate $Q_{\rm tot} (r)$ is much lower than expected
from a ``phenomenological" turbulence model, equation (\ref{eq:Qphen}) with $c_{\rm d} = 1$.
The phenomenological model is based on the assumption that the cascade rates $\epsilon_{\pm}
(r,a_\perp)$ are approximately constant with wavenumber $a_\perp$ in the inertial range,
and that the cascade is everywhere direct. These assumptions are reasonable for a fully developed
turbulent system in a closed box or periodic domain, but they are not appropriate for the extended
corona where the cascade time for the dominant waves is comparable to the wave travel time and
the dissipation is highly non-local. Indeed, we find that the cascade rate $\epsilon_{+}$ varies
significantly with wavenumber in the inertial range. This fact prevents the straightforward
application of the phenomenological formalism. }

We predict that for models with a monotonically decreasing Alfv\'{e}n speed $v_{\rm A} (r)$
a direct cascade rate occurs at all heights. To verify this prediction we constructed
a model (not shown) with a temperature $T_0 (r)$ that decreases monotonically with height.
The model was constructed by setting $C_0 = 0.3$, $C_1 = -2$ and $k=4$ in equation (59) of
paper~I; the coronal base pressure was assumed to be $0.03$ $\rm dyne ~ cm^{-2}$.
In this model the Alfv\'{e}n speed $v_{\rm A} (r)$ decreases monotonically with height.
We find that indeed the cascade rate $\epsilon_{+} > 0$ at all heights for wavenumbers
larger than those of the driver waves ($a_\perp > 15$). However, such a model for
$v_{\rm A} (r)$ is not realistic. In the chromospheric-corona transition region
the temperature $T_0 (r)$ must increase with height, and the density must rapidly decrease,
leading to a local increase in Alfv\'{e}n speed $v_{\rm A} (r)$ with height. In the corona we
expect high Alfv\'{e}n speeds, $v_{\rm A} \sim 1000$ $\rm km ~ s^{-1}$. In the solar wind
the Alfv\'{e}n speed is again relatively low, $v_{\rm A} \sim 30$ $\rm km ~ s^{-1}$ at 1 AU.
Therefore, for a realistic model of the fast solar wind the Alfv\'{e}n speed $v_{\rm A} (r)$
must have a maximum at some height in the corona. Hence, inverse cascade of the kind found
in the present models cannot easily be avoided.

The models presented here and in paper~I differ significantly from other models
in which the driver waves at the coronal base are assumed to have large perpendicular
correlation lengths (10 to 30 Mm) and long correlation times (tens of minutes or longer)
\citep[e.g.,][]{Verdini2009, Perez2013}. Such large length and time scales are comparable
to those of the supergranulation, which is a convective flow pattern observed in the solar
photosphere. Long-period waves are more strongly reflected in the extended corona,
and produce stronger minority waves and higher wave dissipation rates than the present
model. However, as argued in paper~I we do not believe that the supergranulation can play
a significant role in producing the transverse waves that drive the solar wind. The reason
is that supergranular flows have low velocity ($\sim 0.3$ $\rm km ~ s^{-1}$), and the
magnetic elements in the lower atmosphere respond quasi-statically to such weak, slowly
varying flows. The buoyant magnetic flux elements in the photosphere are expected to be
passively advected by these horizontal flows without much amplification of the motions
with increasing height. Therefore, the supergranular flows are expected to produce
velocities of only about 0.3 $\rm km ~ s^{-1}$ in the low corona, not the much larger
velocities assumed by \citet[][]{Verdini2009} and \citet[][]{Perez2013}.

In contrast,
in the present model the waves are asumed to be produced by interactions of magnetic
elements with the solar granulation, which is a different convective pattern with a
typical length scale of 1 Mm and time scale of a few minutes. The magnetic structures
in the lower atmosphere respond dynamically to such short-period disturbances
\citep[e.g.,][]{vanB2014}, producing transverse waves that are amplified from about
1 $\rm km ~ s^{-1}$ in the photosphere to about 30 $\rm km ~ s^{-1}$ in the low corona.
Therefore, in the present model we assume a relatively small correlation length
($\lambda_{\perp,\odot} = 1$ Mm) and a short correlation time ($\tau_{\rm c} = 50$ s).
This produces less wave reflection than in models with long-period waves, and we find
that the minority waves are predominantly of the outward-propagaing ``anomalous" type,
whereas \citet[][]{Verdini2009} find that for long-period driving there is also a
significant component of inward-propagating ``classical" waves. Although we find an
inverse cascade, its effects are limited to length scales smaller than those of the
driver waves, and we do not find a strong tendency for the turbulence to cascade to
larger scales. To distinguish between the different turbulence models will require
further observations of Alfv\'{e}n waves in coronal holes, including their typical
length- and time scales and the relative amplitudes of outward- and inward-propagating
waves. The observations by \citet[][]{Morton2015} are an important step in that
direction.

A number of authors have suggested that density fluctations and coupling between
Alfv\'{e}n- and compressive MHD waves play an important role in the heating of the
fast solar wind \citep[e.g.,][]{Kudoh1999, Moriyasu2004, Suzuki2005, Suzuki2006,
Matsumoto2010}. Here we find that RMHD models with fixed density variations in the
background atmosphere have inverse cascades that limit the rate at which wave energy
can be dissipated. In such models sufficient energy is available at large spatial
scales, but the energy is not efficiently cascaded to small scales where it can be
dissipated. To obtain a more efficient cascade process it may be necessary to go beyond
standard RMHD modeling and include the effects of perpendicular density variations,
which give rise to phase mixing and resonant absorption of the waves \citep[e.g.,][]
{Heyvaerts1983, DeGroof2002, Goossens2011, Goossens2012, Goossens2013, Pascoe2012}.
Future modeling of Alfv\'{e}n wave turbulence in the acceleration region of the fast
wind should take such transverse density variations into account.

\acknowledgements
We thank the referee for useful comments that led us to explore alternative models
for the background atmosphere.
We thank Benjamin Chandran for comments that helped improve the presentation of the
paper. We are most grateful to Alex Voss from the School of Computer Science
at the University of St.~Andrews for his support with the computational work.
This project was supported under contract NNM07AB07C from NASA to the Smithsonian
Astrophysical Observatory (SAO) and SP02H1701R from LMSAL to SAO.

\clearpage

\end{document}